\documentclass[11pt,aps,prc,preprint,showkeys,doublespace,showpacs,
amssymb,byrevtex,superscriptaddress,nofootinbib]{revtex4}  
\usepackage{epsfig,bm,dcolumn}

\begin{document}


\title{Possibility of determining the parity of the pentaquark $\Theta^+$  from
photoproduction near threshold}


\author{Byung Geel Yu}%

\email{bgyu@mail.hangkong.ac.kr}
\affiliation{Department of General studies, 
Hankuk Aviation University, Koyang, 412-791, Korea}
\affiliation{Department of Physics, North Carolina State
University, Raleigh, North Carolina 27695-8202, USA}

\author{Tae Keun Choi}%

\email{tkchoi@dragon.yonsei.ac.kr} \affiliation{Department of
Physics, Yonsei University, Wonju, 220-710, Korea}

\author{Chueng-Ryong Ji}%
\email{crji@unity.ncsu.edu} \affiliation{Department of Physics,
North Carolina State University, Raleigh, North Carolina
27695-8202, USA}

\date{\today}


\begin{abstract}

We discuss the possibility of determining the parity of the
$\Theta^+$ baryon from photoproduction $\gamma N\to K\Theta^+$
process near threshold. We utilize the conservation laws of parity
and angular momentum for the analysis of angular distributions and
spin observables near threshold. Since the discussion is in
essence a partial wave analysis of the production mechanism the
result should be less dependent on the model parameters. Our
analysis shows that the angular distribution and photon
polarization asymmetry for the process of neutron target are
sensitive to the parity of the $\Theta^+$, but not for the case of
proton target. In the case of proton target, the polarization
asymmetries of target and recoiled $\Theta^+$ are preferred for
parity determination.

\end{abstract}


\maketitle

\section{Introduction}

Experimental evidences for the pentaquark
$\Theta^+(1540)$ \cite{theo} have been debated since the first
claim of observing the narrow peak in the LEPS collaboration at
SPring-8 \cite{leps}. In a sequence of the following experiments,
more than 10 experiments reported affirmative results on the
existence of the $\Theta^+$ \cite{saph,posi-ex}. However, many
other experiments, in particular, those performed using $e^+ e^-$
processes and high energy proton beams could not support the
existence of $\Theta^+$ \cite{nega-ex-1}. Recent report from the
CLAS collaboration in the $g_{11}$ experiment \cite{aps05} is
especially disturbing since it refutes the earlier SAPHIR
experiment \cite{saph}. Nevertheless, the existence of the
pentaquark $\Theta^+$ has not yet been ruled out \cite{kali,sp8}
due to the limited experimental factors such as low statistics,
uncertainty in the background effect, and the specific cuts in
angle for data analysis. We think that more refined experiments
are crucial to resolve this debate and further analyses on the
$\Theta^+$ system would be worthwhile till the debate is resolved.

In theoretical side, it is crucial to determine the parity of the
$\Theta^+$ because the parity is a decisive quantum number to
understand its substructure with strangeness $S=+1$
\cite{posi,nega}.
In Ref. \cite{thom} and in subsequent works \cite{hanh1,hanh2},
the conservation laws were applied to give strong constraints on the
parity and angular momentum of the initial $pp$ state in the
polarized process $pp\to\Sigma^+\Theta^+$ near threshold. These
works claimed that the parity of the $\Theta^+$ can be clearly
determined by the Fermi statistics of a two-nucleon system. Also,
Ref. \cite{naka1} suggested that the parity of the $\Theta^+$ can
be determined model-independently by observing the polarization
asymmetries based on the reflection symmetry in the polarized
process of $\gamma N\to K\Theta^+$. In addition, Refs.
\cite{zhao01,reka,zhao} proposed to use a polarized photon beam
for the process $\gamma N\to K\Theta^+$ to determine the
$\Theta^+$ parity model-independently. These works emphasize that
the polarization observables are important tools to determine the
$\Theta^+$ parity.

In this work, we discuss a possibility that the conservation laws
of parity and angular momentum for the $unpolarized$ observable
can also be exploited to determine the parity of the $\Theta^+$ in
the photoproduction process, if the photon energy lies near
threshold. We present in detail the role of parity and angular
momentum conservation on the process $\gamma N\to K\Theta^+$ and
show that it is possible to describe the amplitude of the process
near threshold from first principles with the threshold
kinematics.
As a result, the two assumed states of the $\Theta^+$
corresponding to each parity can be distinguished from each other
by analyzing simply the angular distribution of the unpolarized
photoproduction process. In the same line of reasoning we can also
extend this sort of analysis to the polarized process and make a
prediction for the polarization asymmetries of the photon beam (or
of the target) and the recoiled $\Theta^+$. Similar type of
analysis for the photoproduction process has already been applied
to pion photo and electroproduction processes near threshold
\cite{mainz,donn,hans}. The analogy of the charge coupling
structures between the processes $\gamma n\to K^-\Theta^+$($\gamma
p\to \bar{K}^0\Theta^+$) and $\gamma n\to \pi^- p$($\gamma p\to
\pi^0 p$) is useful for our analysis.

For explicit quantitative predictions, we present numerical
results of the angular distribution and single polarization asymmetries
using the hadron model in Ref. \cite{bgyu}.
The model dependence in this case, however, can be minimized near
threshold because the conservation laws can be implemented only to
the lower states of angular momentum available near threshold.

In Section 2, we present our reasoning for the difference in the
angular distribution depending on the parity of the $\Theta^+$
rather than total cross section. The conservation laws of parity
and angular momentum are applied for identifying the leading
multipole in the photoproduction of the positive and negative
parity of the $\Theta^+$. The CGLN amplitudes based on the
threshold kinematics are also used for discussing consistency of
our description. Numerical estimations follow for illustration.
Section 3 is devoted to the analysis of the angular distribution
and single polarization observables with respect to the parity of
$\Theta^+$. Summary and discussion follow in Section 4.

\section{ Conservation laws and threshold
kinematics}

Before we proceed to analyze the angular distribution of $\gamma
N\to K\Theta^+$ based on the conservation laws of the parity and
angular momentum, we first start with a summary of our reasoning
why the total cross sections alone may not be so effective in
determining the $\Theta^+$ parity.

\begin{figure}[tbh]
\centering%
\includegraphics[width=10cm]{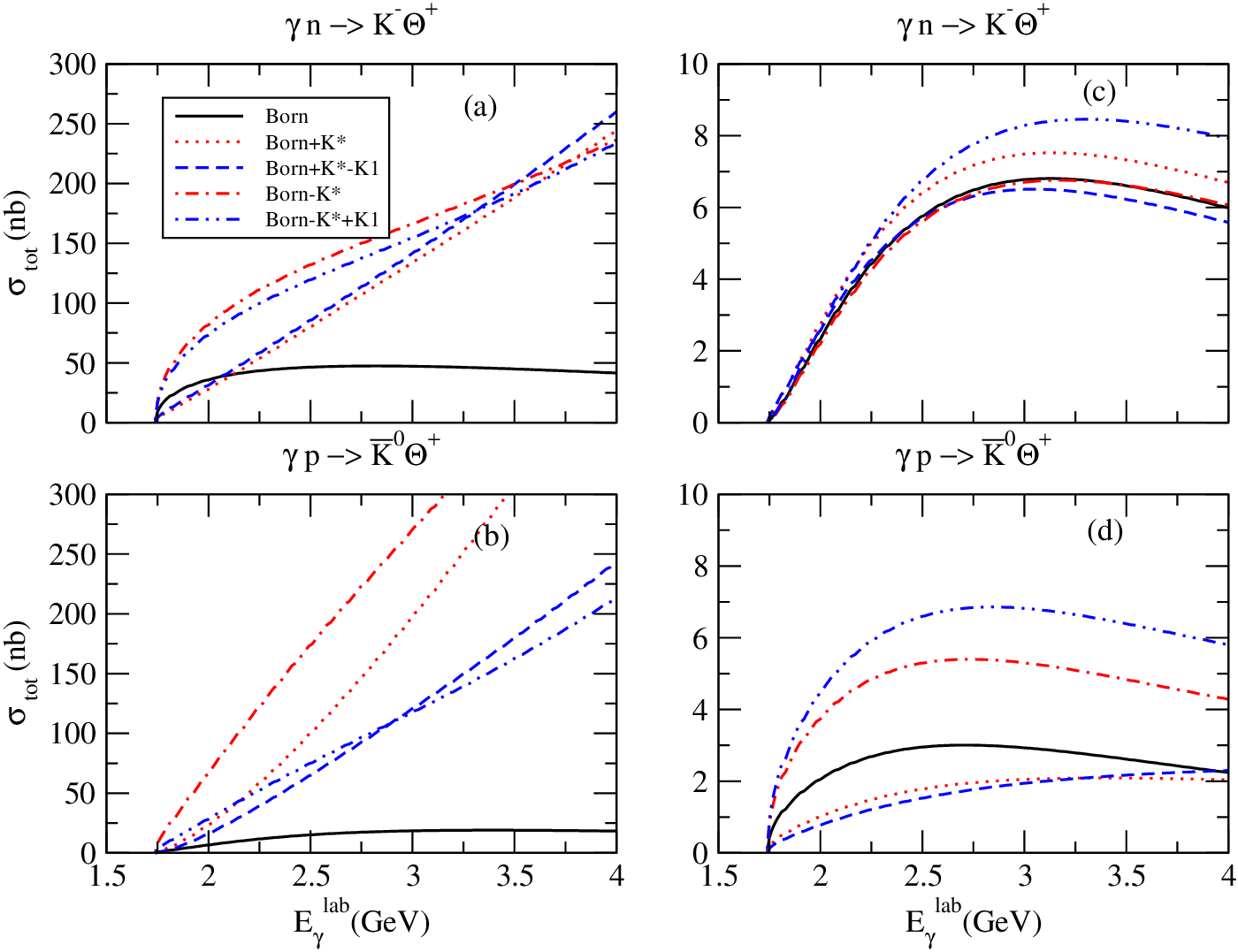}
\caption{ Total cross sections of the $\gamma N \to K \Theta^+$ in
the PV coupling scheme. Panels (a) and (b) ((c) and (d)) are for
$\Theta^+(\frac{1}{2}^+)$ ($\Theta^+(\frac{1}{2}^-)$).
The solid lines are the contributions
of the Born terms with $g_{KN\Theta}=0.984(0.137)$ and
$g_{K^*N\Theta}=0$, $g_{K_1 N\Theta}=0$. The dotted lines are the
sum of the Born terms and $K^*$ contribution with $g_{K^*N\Theta}=
1.704(0.08)$. The dot-dashed lines are the sum of the Born terms
and $K^*$ with $g_{K^*N\Theta}=-1.704(-0.08)$. The dashed lines
are the sum of the Born terms, $K^*$ and $K_1$ contributions with
$g_{K^*N\Theta}= 1.704$, $g_{K_1 N\Theta}= -0.09(-0.138)$ for
$\gamma n\to K^-\Theta^+(\gamma p\to \bar{K}^0\Theta^+)$ for
$\Theta^+(\frac{1}{2}^+)$ and $g_{K^*N\Theta}= 0.08$, $g_{K_1
N\Theta}= -0.048(-0.074)$ for $\gamma n\to K^-\Theta^+(\gamma p\to
\bar{K}^0\Theta^+)$ for $\Theta^+(\frac{1}{2}^-)$. The
dot-dot-dashed lines are the sum of the Born terms, $K^*$
and $K_1$ contributions with $g_{K^*N\Theta}= -1.704$, $g_{K_1 N\Theta}=
+0.09(+0.138)$ for $\gamma n\to K^-\Theta^+(\gamma p\to
\bar{K}^0\Theta^+)$ for $\Theta^+(\frac{1}{2}^+)$ and
$g_{K^*N\Theta}= -0.08$, $g_{K_1 N\Theta}= +0.048(+0.074)$ for
$\gamma n\to K^-\Theta^+(\gamma p\to \bar{K}^0\Theta^+)$ for
$\Theta^+(\frac{1}{2}^-)$. } \label{fig:fig00}
\end{figure}

Fig. \ref{fig:fig00} shows the cross sections for $\gamma N\to
K\Theta^+$ obtained from our previous work \cite{bgyu} with the
coupling constant $g_{KN\Theta}$ taken from the width
$\Gamma_{\Theta}=1$ MeV for both parities \cite{kn}. Following
Refs. \cite{dude,carl1,carl2}, the coupling constant $K^*$ and
subsequently that of $K_1$ have been updated as
$g_{K^*N\Theta}=\sqrt{3}g_{KN\Theta}$ for the positive parity, and
$g_{K^*N\Theta}=\frac{1}{\sqrt{3}}g_{KN\Theta}$ for the negative
parity of the $\Theta^+$, respectively. The tensor coupling
constants of the $K^*$ and $K_1$ are also neglected to avoid
unnecessary complication in our analysis near threshold. The
coupling constants used in the calculation are listed in Table
\ref{cc}.

While the reanalysis of the SAPHIR data
lowered the magnitude of the total cross section of $\gamma p\to
{\bar K^0}\Theta^+$ from initially 300 nb to around 50 nb, the
most recent analysis from the CLAS collaboration reported even
further reduction with an upper bound of $1 \sim 4$ nb at most
\cite{aps05}. As can be seen in Fig. \ref{fig:fig00}, our total
cross section of the $\gamma p\to\bar{K}^0\Theta^+$ process for
the case of negative parity $\Theta^+$(Fig. \ref{fig:fig00}(d))
seems now to agree with the recent CLAS analysis. However, our total
cross section can change to agree with the SAPHIR data (50 nb)
if the decay width $\Gamma_{\Theta}=5$ MeV is taken as in
the case of earlier analysis. Moreover, the CLAS measurements
and the SPring8 measurements \cite{sp8} have been exclusive
to each other in the
sense that the SPring8 measurements were limited only to the
forward direction of the produced particles but the CLAS
measurements excluded this forward direction entirely. Therefore,
although our results for the photoproduction of the negative
parity $\Theta^+$ using the proton target are consistent with the
most recent CLAS data, a definite conclusion from the
total cross section cannot be made until the improvement is made
to match kinematic regions between the two facilities.

It is, thus, necessary to go beyond the total cross sections and
investigate angular distributions and possibly other
observables for discriminating the $\Theta^+$ parity
\cite{bgyu}. Although the hadron model involves several unknown
parameters, we expect that such model dependence can be minimized
in the analysis near threshold because only the lower angular
momentum states are available and thus it is rather easy to
implement the first principle conservation laws \cite{thom}.
\begin{table}
\caption{\label{cc} Coupling constants used in the calculation for
$\gamma N \to K\Theta^+$. $\Gamma_{\Theta^+\to NK}=$ 1 MeV taken
for $g_{KN\Theta}$ of the $\Theta^+(\frac{1}{2}^{\pm})$
\cite{kn}. Superscripts a, b, c and d in the last column of References
denote $^a$the assumption \cite{dude,carl1,carl2},
$^b$PDG \cite{pdg}, $^c$$\gamma p\to K^+\Lambda$ \cite{crji} with
assumption $\frac{G_{K^{*}N\Theta}}{G_{K_{1}N\Theta}}\simeq
\frac{G_{K^{*}N\Lambda(1116)}}{G_{K_{1}N\Lambda(1116)}}$ for
$\Theta^+(\frac{1}{2}^{+})$ and
$\frac{G_{K^{*}N\Theta}}{G_{K_{1}N\Theta}}\simeq
\frac{G_{K^{*}N\Lambda(1405)}}{G_{K_{1}N\Lambda(1405)}}$ for
$\Theta^+(\frac{1}{2}^{-})$ with $G_{K^{*}(K_1)N
Y}=g_{K^*(K_1)K\gamma}g_{K^*(K_1)NY}$,  and $^d$$\Gamma_{K_1
K\rho}=37.8$ MeV with Vector Meson Dominance $g_{K_1
K\gamma}=\frac{e}{f_{\rho}}g_{K_1 K \rho}$ \cite{hagl}. Values in
the parenthesis denote $\gamma n\to K^-\Theta^+$($\gamma p\to
\bar{K}^0\Theta^+$) respectively.}
\begin{ruledtabular}\label{cc}
\begin{tabular}{cccl}
               & Positive parity & Negative parity & References \\
\hline\hline $g_{KN\Theta}$ & 0.984 & 0.137 & $\Gamma_{\Theta^{+}\to
NK}$ \\
\hline $g_{K^*N\Theta}$ & $\pm$1.704   & $\pm$0.08   &
$\frac{g_{K^*N\Theta}}{g_{KN\Theta}}=\sqrt{3}, (\frac{1}{\sqrt{3}})^a$ \\ $g_{K^*K\gamma}$ &
0.254(0.388)& 0.254(0.388)& $\Gamma_{K^{*}\to K\gamma}$$^b$ \\
\hline $g_{K_1 N\Theta}$ & $\mp$ 0.09(0.138)&                &
$\frac{G_{K^{*}N\Lambda(1116)}}{G_{K_{1}N\Lambda(1116)}}\simeq
-8^c$ \\ $               $ &                & $\mp$0.048(0.074) &
$\frac{G_{K^{*}N\Lambda(1405)}}{G_{K_{1}N\Lambda(1405)}}\simeq
-0.7^c$ \\ $g_{K_1 K\gamma}$ & 0.6 & 0.6 & $g_{K_1 K \rho}$$^d$ \\
\end{tabular}
\end{ruledtabular}
\end{table}
\vspace{0.5cm}
\begin{table}
\caption{\label{cd} Multipole states for $\gamma N \to K\Theta^+$.
For the transverse photon states, $L\geq 1$. The angular momentum
$J=|l\pm \frac{1}{2}|=|L\pm \frac{1}{2}|$. From the parity
conservation  for $\Theta^+(\frac{1}{2}^+)$ E$L$ :
$(-1)^L=(-1)^{l+1}$, M$L:(-1)^{L+1}=(-1)^{l+1}$. For
$\Theta^+(\frac{1}{2}^-)$ E$L:(-1)^{L}=(-1)^{l}$,
M$L:(-1)^{L+1}=(-1)^{l}$.}
\begin{ruledtabular}
\begin{tabular}{c|cccc|cccc}
    & \multicolumn{4}{c|}{Positive parity} &\multicolumn{4}{c}{Negative parity}\\
\hline\hline $L$&E$L$(M$L$)& $J^P$          &$l$& multipoles
&E$L$(M$L$)&$J^P$           &$l$ &multipoles\\ \hline       $1$&
E1       &$\frac{1}{2}^-$ & 0 &$E_{0+}$    & M1
&$\frac{1}{2}^+$ & 0 & $M_{0+}$\\ $1$ &  M1   & $\frac{1}{2}^+$ &
1 &$M_{1-}$ & E1& $\frac{1}{2}^-$ & 1 &$E_{1-}$  \\ \hline $1$ &
M1   & $\frac{3}{2}^+$ & 1 &$M_{1+}$ & E1&      $\frac{3}{2}^-$ &
1 &$E_{1+}$  \\ $2$ &  E2   & $\frac{3}{2}^+$ & 1 &$E_{1+}$ & M2&
$\frac{3}{2}^-$ & 1 &$M_{1+}$\\
\end{tabular}
\end{ruledtabular}
\end{table}

In the threshold region it is legitimate to assume that the
angular momentum of kaon is $l=0$ or $1$ in the final $K\Theta^+$
state, because a typical hadronic scale $R$ is one fermi and
$|\bm{q}|R\approx \sqrt{l(l+1)}\leq 2$ up to a few
hundred MeV/c of the kaon momentum $|\bm{q}|$ in the center of
mass(CM) frame.
The angular momentum of the final state is therefore either
$J=\frac{1}{2}$ or $\frac{3}{2}$. The parity of the final state is
given by $(-1)^{l+1}$ for the positive parity of the
$\Theta^+(\frac{1}{2}^+)$, and by $(-1)^l$ for the negative parity
of the $\Theta^+(\frac{1}{2}^-)$. The angular momentum of the
initial $\gamma N$ state is given by $J=|L\pm \frac{1}{2}|$, where
$L$ is the total orbital angular momentum of the photon
\cite{mainz}. For transverse photon, the parity of the initial
state can have either $(-1)^L$ for electric, or $(-1)^{L+1}$ for
magnetic states of the photon \cite{mainz,donn}.
Then, from the conservation of parity and angular momentum, only
s- and p-waves are allowed, as summarized in Table \ref{cd}. Near
threshold, however, the electric transition must be dominant over
the magnetic transition. Also the lower state of $J$(i.e.,
$J=\frac{1}{2}$) must be energetically more accessible than the
higher state of $J$(i.e., $J=\frac{3}{2}$). Therefore, the
dominant transitions near threshold are $E_{0+}$ for the
$\Theta^+(\frac{1}{2}^+)$, and $E_{1-}$ for the
$\Theta^+(\frac{1}{2}^-)$, respectively. This would make a clear
distinction between the two angular distributions for the $\gamma
N\to K\Theta^+$ process, depending on the $\Theta^+$ parity. In
this work we show that this expectation generally holds for the
case of $\gamma n\to K^-\Theta^+$, while the case of $\gamma p\to
\bar{K}^0\Theta^+$ does not share the same features due to the
absence of the charge coupling of photon to neutral $\bar{K}^0$
meson.

\subsection{Consistency with CGLN Amplitudes}

To check the consistency of our description, let us now expand the
photoproduction current in terms of CGLN amplitudes \cite{cgln}.
In the case of $\Theta^+(\frac{1}{2}^+)$, the CGLN expansion is
given by
\begin{eqnarray}\label{cgln+}
\bm{J}^+\cdot\bm{\hat\epsilon}_{\lambda}
=
F^+_1\,i\bm{\sigma}\cdot\bm{\hat\epsilon}_{\lambda}
+F^+_2\,\bm{\sigma}\cdot\bm{\hat{q}}\,
\bm{\sigma}\cdot({\bm{\hat{k}}\times\bm{\hat\epsilon}_{\lambda}})
+F^+_3\,i\bm{\sigma}\cdot\bm{\hat{k}}\,\bm{\hat{q}}\cdot\bm{\hat\epsilon}_{\lambda}
+F^+_4\,i\bm{\sigma}\cdot\bm{\hat{q}}\,\bm{\hat{q}}\cdot\bm{\hat\epsilon}_{\lambda},
\end{eqnarray}
where $\bm{\hat{k}}$ and $\bm{\hat{q}}$ are unit vectors of photon
and kaon three momenta, respectively, and
$\bm{\hat\epsilon}_{\lambda}$ is the photon polarization vector
with the polarization $\lambda$. Likewise, the expansion of
photoproduction current for the $\Theta^+(\frac{1}{2}^-)$ is given
by \cite{taba}
\begin{eqnarray}\label{cgln-}
\bm{J}^-\cdot\bm{\hat\epsilon}_{\lambda}
=
F^-_1\,i\bm{\sigma}\cdot(\bm{\hat
k}\times\bm{\hat\epsilon}_{\lambda})
+F^-_2\,\bm{\sigma}\cdot\bm{\hat{q}}\,\bm{\sigma}\cdot\bm{\hat\epsilon}_{\lambda}
+F^-_3\,i\bm{\sigma}\cdot(\bm{\hat{q}}\times
\bm{\hat{k}})\,\bm{\hat{q}}\cdot\bm{\hat\epsilon}_{\lambda}
+F^-_4\,\bm{\hat{q}}\cdot\bm{\hat\epsilon}_{\lambda}\,.
\end{eqnarray}
Here and in what follows, the superscripts $\pm$ stand for the two
possible parities of the $\Theta^+$.
The CGLN amplitudes in Eqs. (\ref{cgln+}) and (\ref{cgln-}) are
given by
\begin{eqnarray}\label{cglnftn+}
&&F_1^{\pm}=\pm\frac{|\bm{k}|}{4\pi}N_{+} \biggl[
A^{\pm}_1\mp\frac{1}{2}(W\pm M)A^{\pm}_3 \mp\frac{p^{\prime} \cdot
k}{W\mp M} A^{\pm}_4 \biggr], \nonumber\\
&&F_2^{\pm}=\pm\frac{|\bm{k}|}{4\pi}N_{-}
\biggl[ A_1^{\pm}\pm\frac{1}{2}(W\mp M)A^{\pm}_3 \pm\frac{p^{\prime}
\cdot k}{W\pm M} A^{\pm}_4 \biggr], \nonumber\\
&&F_3^{\pm}=\mp\frac{|\bm{k}||\bm{q}|}{4\pi}N_{\pm} \biggl[(W-M)
A^{\pm}_2 -A^{\pm}_4 \biggr], \nonumber\\
&&F_4^{\pm}=\pm\frac{|\bm{k}||\bm{q}|}{4\pi}N_{\mp} \Biggl[(W+M)
A^{\pm}_2 +A^{\pm}_4
-\frac{(W+M)\bm{k}\cdot\bm{q}}{(E+M)(E_{\Theta}+M_{\Theta})}
\left(A^{-}_2-\frac{A^{-}_4}{W-M}\right)\Biggr],
\end{eqnarray}
with the normalization constants
$N_{\pm}=\sqrt{\frac{E_{\Theta}\pm M_{\Theta}}{2W}}$\,\,.

We now note that, regardless of any model descriptions for
$A_i^{\pm}$, essentially the knowledge of the kinematics near
threshold enables us to determine which amplitude in the CGLN
expansion of Eqs.(\ref{cgln+}) and (\ref{cgln-}) gives the leading
contribution to the currents ${\bm J}^{\pm}$. First,  the
amplitudes $F^+_{2}$ and $F^+_{4}$ in Eq.(\ref{cgln+}) are
substantially suppressed near threshold due to the kinematic
constant $N_{-}$. Therefore, the current given by Eq.(\ref{cgln+})
near threshold can be written as
$\bm{J}^+\cdot\bm{\hat\epsilon}_{\lambda}$ $ \simeq
F^+_1\,i\bm{\sigma}\cdot\bm{\hat\epsilon}_{\lambda}
+F^+_3\,i\bm{\sigma}\cdot\bm{\hat{k}}\,\bm{\hat{q}}
\cdot\bm{\hat\epsilon}_{\lambda}$. Similarly, the amplitudes
$F^-_2$ and $F^-_3$ are negligible near threshold by the same
reason and thus the current in Eq.(\ref{cgln-}) can be given by
$\bm{J}^-\cdot\bm{\hat\epsilon}_{\lambda}$ $ \simeq
F^-_1\,i\bm{\sigma}\cdot(\bm{\hat
k}\times\bm{\hat\epsilon}_{\lambda})
+F^-_4\,\bm{\hat{q}}\cdot\bm{\hat\epsilon}_{\lambda}$.

For the case of the $\Theta^+(\frac{1}{2}^+)$, it is apparent that
near threshold the $F_1^+\,i\bm{\sigma}\cdot\bm{\hat\epsilon}_{\lambda}$ is dominant over
the $F_3^+\,i\bm{\sigma}\cdot\bm{\hat{k}}\,\bm{\hat{q}}
\cdot\bm{\hat\epsilon}_{\lambda}$
due to the dominance of the electric transition over the magnetic
transition near threshold, viz., $L=1$ in Table \ref{cd}. Note
that the $i\bm{\sigma}\cdot\bm{\hat\epsilon}_{\lambda}$ term
governs the s-wave multipole of outgoing kaon via electric
transition while the $i\bm{\sigma}\cdot\bm{\hat{k}}\,\bm{\hat{q}}
\cdot\bm{\hat\epsilon}_{\lambda}$ term does the p-wave multipole
via magnetic transition. For the $\Theta^+(\frac{1}{2}^-)$,
however, there is a turnover between the electric and magnetic
transitions according to the change of the positive parity of the
$\Theta^+$ to the negative one, as shown in Table \ref{cd}. The
$i\bm{\sigma}\cdot(\bm{\bm{\hat k}\times\hat\epsilon}_{\lambda})$
term governs the s-wave multipole via magnetic transition, while
the $\bm{\hat{q}}\cdot\bm{\hat\epsilon}_{\lambda}$ does the p-wave
one via electric transition. Thus, in identifying the leading
multipole in the CGLN amplitudes for the case of
$\Theta^+(\frac{1}{2}^-)$, there is a subtle difference between
the $\gamma n\to K^-\Theta^+$ and $\gamma p\to \bar{K}^0\Theta^+$
processes\footnote{It is, of course, true that at threshold
$E_{\gamma}=1.75$ GeV where $|\bm{q}|=0$ exactly only the s-wave
multipole $F_1^{\pm}$ survives in any cases \cite{reka,zhao01}.
Note that, however, the energy region we are concerned with is the
region, slightly above threshold, $E_{\gamma}=1.8$ GeV which
allows the nonvanishing kaon momentum. Therefore in this region
the magnetic dipole transition by small kaon momentum is dominated
by the electric one. }.
Although one expects that $F_4^-
\bm{\hat{q}}\cdot\bm{\hat\epsilon}_{\lambda}$ be dominant over
$F_1^- i\bm{\sigma}\cdot(\bm{\bm{\hat
k}\times\hat\epsilon}_{\lambda})$ on account of the dominance of
the electric transition over the magnetic transition near
threshold, this is fulfilled only in the $\gamma n\to K^-\Theta^+$
of the negative parity $\Theta^+$ but not in the $\gamma p\to
\bar{K}^0\Theta^+$. In case of the latter process, both the s-wave
multipoles, $F_1^+\,i\bm{\sigma}\cdot\bm{\hat\epsilon}_{\lambda}$
and $F_1^- i\bm{\sigma}\cdot(\bm{\bm{\hat
k}\times\hat\epsilon}_{\lambda})$, are the leading ones to both
cases of the $\Theta^+$ parities near threshold due to the absence
of the t-channel kaon exchange from the neutralness of
$\bar{K}^0$. The electric transition governed by the
$F_4^-\bm{\hat{q}}\cdot\bm{\hat\epsilon}_{\lambda}$ term, thus,
cannot play a dominant role in the case of the $\gamma p\to
\bar{K}^0\Theta^+$ process. On the other hand, one also needs to
take into account the fact that
$F_1^-i\bm{\sigma}\cdot(\bm{\bm{\hat
k}\times\hat\epsilon}_{\lambda})$ is suppressed and becomes
smaller than $F_4^- \bm{\hat{q}}\cdot\bm{\hat\epsilon}_{\lambda}$
near threshold in the case of $\gamma n\to K^-\Theta^+$
\cite{taba, marc}. Such a suppression of magnetic transition near
threshold in the case of $F_1^-i\bm{\sigma}\cdot(\bm{\bm{\hat
k}\times\hat\epsilon}_{\lambda})$ can be checked by its kinematic
factor of $\frac{|\bm{k}|}{E+M}$ \footnote{The nonrelativistic
reduction of the vertex $KN\Theta^+(\frac{1}{2}^-)$ in the CM
frame yields, $$
V_{KN\Theta}=\frac{eg}{M_{\Theta}-M}\bar{u}(p^{\prime})\,
/\kern-6pt{\epsilon}u(p) =\frac{eg}{M_{\Theta}-M}NN^{\prime}
\frac{|\bm{k}|}{E+M}\chi^{\dagger} i\bm{\sigma}\cdot(\bm{\bm{\hat
k}\times\hat\epsilon}_{\lambda})\chi + \cdots $$ with
$N(N^{\prime})$ the normalization constant of initial(final) Dirac
spinor. At the photon energy $E_\gamma$=1.8 GeV in the lab. frame,
the kinematic factor $\frac{k}{E+M}\simeq 0.37$ in the CM frame.},
which makes $F_1^-$ by an order of magnitude smaller than $F_1^+$,
even compared with the positive parity in the same process.

Summarizing above, we thus identify the leading contribution to
the currents in Eqs.(\ref{cgln+}) and (\ref{cgln-}) as,
\begin{eqnarray}\label{gamma n}
&&\bm{J}^+\cdot\bm{\hat\epsilon}_{\lambda} \simeq
F^+_1\,i\bm{\sigma}\cdot\bm{\hat\epsilon}_{\lambda}\,,\nonumber\\
&&\bm{J}^-\cdot\bm{\hat\epsilon}_{\lambda} \simeq
F^-_4\,\bm{\hat{q}}\cdot\bm{\hat\epsilon}_{\lambda}\,,
\end{eqnarray}
for $\gamma n\to K^-\Theta^+$ \cite{zhao} and
\begin{eqnarray}\label{gamma p}
&&\bm{J}^+\cdot\bm{\hat\epsilon}_{\lambda} \simeq
F^+_1\,i\bm{\sigma}\cdot\bm{\hat\epsilon}_{\lambda}\,,\nonumber\\
&&\bm{J}^-\cdot\bm{\hat\epsilon}_{\lambda} \simeq
F^-_1\,i\bm{\sigma}\cdot(\bm{\hat
k}\times\bm{\hat\epsilon}_{\lambda})\,,
\end{eqnarray}
for $\gamma p\to \bar{K}^0\Theta^+$ \cite{reka}, respectively near
threshold. These observations are consistent with our previous
discussion based on the conservation laws listed in Table
\ref{cd}.

\subsection{Numerical Check}

For numerical illustration, we estimate the dependence of CGLN
amplitudes $F_i^{\pm}$ on the energy and angle using the hadron
model in Ref. \cite{bgyu} with coupling constants given in Table
\ref{cc} and the invariant amplitudes $A^{\pm}_i$ of the
pseudovector(PV) coupling scheme given by
\begin{eqnarray}\label{born+}
A^{\pm}_{1}&=&e g_{KN\Theta}\left[
 \frac{F_1(s)(\frac{1}{2}(1+\tau_3)+\kappa_N)}{s-M^2}
 +\frac{F_2(u)(1\pm \kappa_{\Theta})}{u-M^2_{\Theta}+i\Gamma_{\Theta}M_{\Theta}}
\right] \pm \frac{eg_{KN\Theta}}{M_{\Theta}\pm
M}\left(F_1(s)\frac{\kappa_N}{2M}
+F_2(u)\frac{\kappa_{\Theta}}{2M_{\Theta}}\right)\nonumber\\&&
+\frac{G_{V}^{K^*}}{m}F_3(t)(M\pm
M_{\Theta})\Delta_{K^*},\nonumber\\
A^{\pm}_{2}&=&\frac{-2eg_{KN\Theta}
\,\widehat{F}}{X(u-M^2_{\Theta}+i\Gamma_{\Theta}M_{\Theta})}\,\tau_3,\nonumber\\
A^{\pm}_{3}&=&\frac{eg_{KN\Theta}\,F_1(s)}{s-M^2}\frac{\kappa_N}{M}
+\frac{G_{V}^{K^*}}{m}F_3(t)\Delta_{K^*} \pm
\frac{G_{V}^{K_1}}{m}F_3(t)\Delta_{K_1},\nonumber\\
A^{\pm}_{4}&=&\frac{eg_{KN\Theta}\,F_2(u)}{u-M^2_{\Theta}+i\Gamma_{\Theta}M_{\Theta}}
\frac{\kappa_{\Theta}}{M_{\Theta}}
+\frac{G_{V}^{K^*}}{m}F_3(t)\Delta_{K^*}
\mp\frac{G_{V}^{K_1}}{m}F_3(t)\Delta_{K_1}\,.
\end{eqnarray}
Here $F_{1}(s)$ and $F_{2}(u)$ are hadronic form factors in the s-
and u-channel, respectively, $\widehat{F}$ is a subtraction
function to restore gauge invariance of the process and $F_3(t)$
is the vertex form factor for the t-channel meson exchanges
including $K^*$ and $K_1$. In Eq. (\ref{born+}), the anomalous
magnetic moments of proton and neutron are $\kappa_p=1.79$ and
$\kappa_n=-1.91$, respectively. Also the factor $X$ in Eq.
(\ref{born+}) is given by $X=(s-M^2)$ for $\gamma p\to
\bar{K}^0\Theta^+$, and $X=(t-m_K^2)$ for $\gamma n\to
K^-\Theta^+$, respectively. Since our calculations show that the
results near threshold are not quite sensitive to the value of the
$\Theta^+$ anomalous magnetic moment, we take $\kappa_{\Theta}=0$
for simplicity.
For $K^*$ and $K_1$ exchanges in the t-channel,  the coupling
constants are $G_{V}^{K^*}=g_{\gamma KK^*}g_{K^*N\Theta}$ and
$G_{V}^{K_1}=g_{\gamma KK_1}g_{K_1 N\Theta}$ with a parameter of
mass dimension, $m$ and the propagators,
$\Delta_{K^*(K_1)}=[t-m_{K^*(K_1)}^2 +
i\Gamma_{K^*(K_1)}m_{K^*(K_1)}]^{-1}$ \cite{bgyu}.

\begin{figure}[tbh]
\centering
\includegraphics[width=10cm]{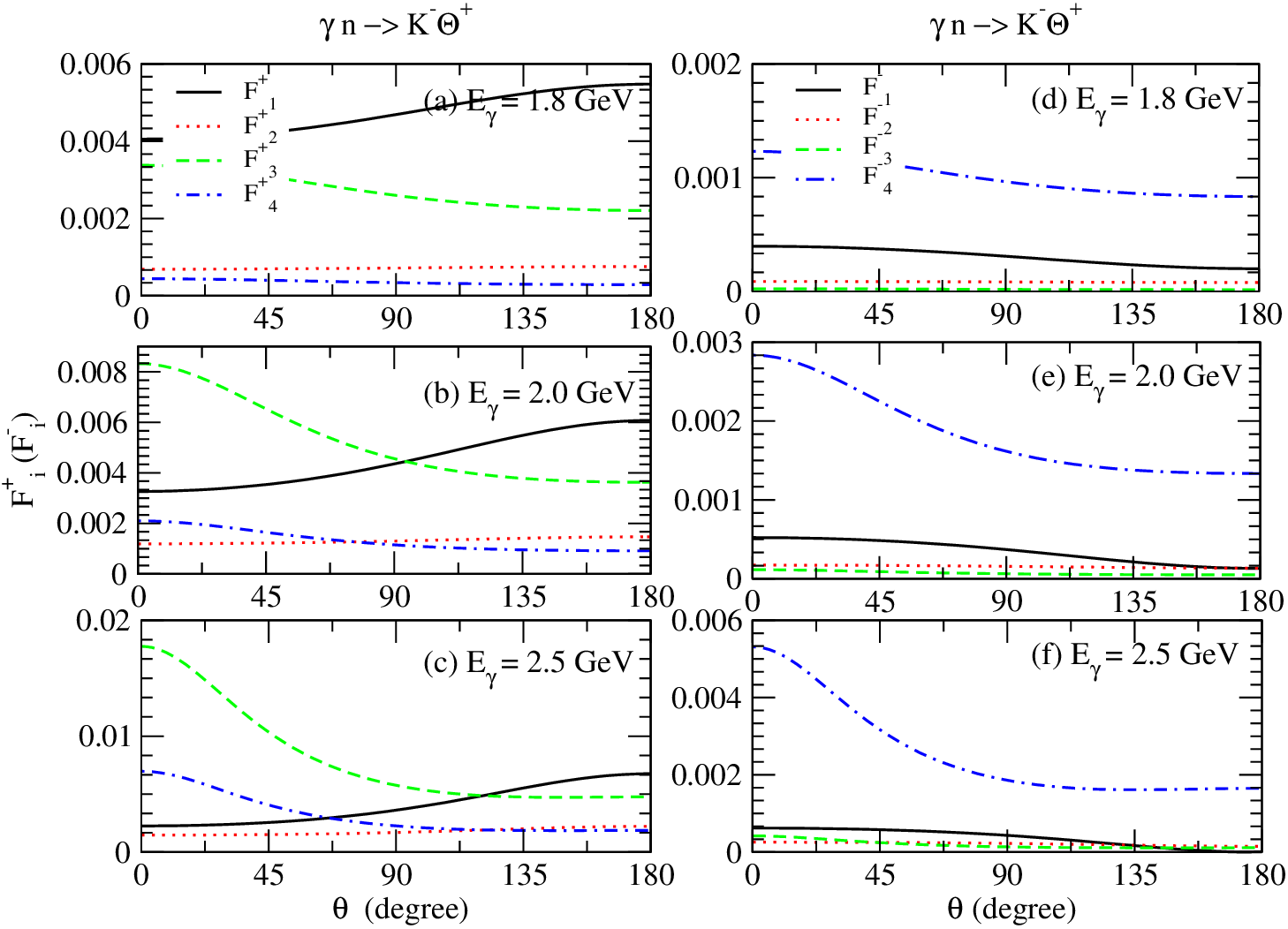}
\caption{(Color online) Angle and energy dependence of the CGLN
amplitudes $F^{+}_i$'s for the $\Theta^+(\frac{1}{2}^+)$((a),(b),(c)) and
$F^{-}_i$'s for the $\Theta^+(\frac{1}{2}^-)$((d),(e),(f)) in the $\gamma n\to
K^-\Theta^+$.  }
\label{fig:fig03}
\end{figure}

\begin{figure}[tbh]
\centering%
\includegraphics[width=10cm]{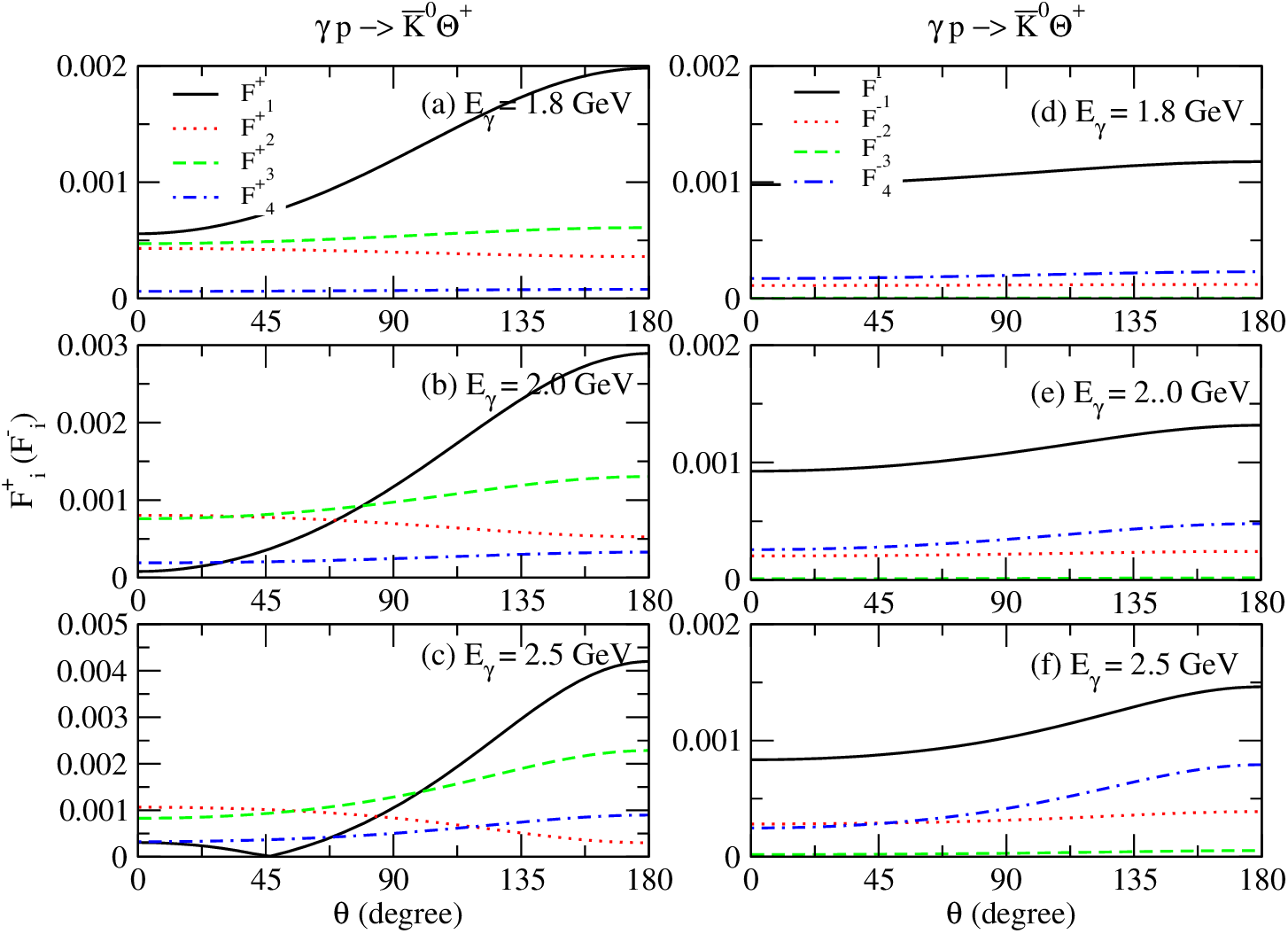}
\caption{(Color online) Angle and energy dependence of the amplitudes
$F^{+}_i$'s for the $\Theta^+(\frac{1}{2}^+)$((a),(b),(c)) and $F^-_i$ for
the $\Theta^+(\frac{1}{2}^-)$((d),(e),(f)) in the $\gamma p \to
\bar{K}^0 \Theta^+$. }
\label{fig:fig04}
\end{figure}

Figs. \ref{fig:fig03} and \ref{fig:fig04} show the results
of the CGLN amplitudes in Eq.(\ref{cglnftn+}) with
$g_{K*N\Theta}=0$ and $g_{K_1N\Theta}=0$.
Addition of the $K^*$ and $K_1$ contributions to
Eq.(\ref{cglnftn+}) cannot alter significantly the leading order
amplitude among $F_i^{\pm}$'s as shown in the figures at
$E_\gamma=1.8$ GeV. In Fig. \ref{fig:fig03}, the features of
$\gamma n\to K^-\Theta^+$ at $E_\gamma=1.8$ GeV are represented as
a dominance of
$F_{1}^{+}$ 
in the $\Theta^+(\frac{1}{2}^+)$ and $F_{4}^{-}$
in the $\Theta^+(\frac{1}{2}^-)$, respectively(see the similar
conclusion on the $\gamma n\to K^-\Theta^+$ process  in Ref.
\cite{zhao01}). Note that Fig. \ref{fig:fig03} reproduces the
suppression of $F_{1}^-$ by an order of magnitude in comparison
with $F_{1}^+$. In Fig. \ref{fig:fig04} one can see that $F_1^+$
and $F_1^-$ are the leading amplitudes at $E_{\gamma}=1.8$ GeV to
the $\gamma p\to \bar{K}^0\Theta^+$ process. Therefore, in the
threshold region, these  figures seem to support our discussions
given above, although estimated by the model-dependent
calculation.

Before closing this section it is worth noting that Ref.
\cite{reka} also discussed contributions of partial waves in
determining the $\Theta^+$ parity from the $\gamma N\to K\Theta^+$
process.

\vspace{0.3cm}
\vspace{1cm}
However, since they did not consider a suppression by a
kinematical factor such as $\frac{k}{E+M} \approx 0.37$ which is
important in the analysis of the $\gamma n\to K^-\Theta^+$ process
near threshold, they started with the s-wave amplitude common to
both parities of the $\Theta^+$. Thus, they need to observe spin
observables more than single and double polarizations instead of
the unpolarized ones, or to use a depolarization tensor component
for relating the spin observables with unpolarized cross sections
for the model-independent analysis. This makes their results
different from ours in the case of $\gamma n\to K^-\Theta^+$.

\section{parity determination from Differential cross section and spin observables}
\subsection{Angular Distribution and Photon polarization asymmetry}

We now analyze the angular distribution. The angular distribution
is defined by
\begin{equation}\label{diff}
\frac{d\sigma}{d\Omega}^{\pm} =
\frac{1}{4}\sum_{s,s'}\frac{|\bm{q}|}{|\bm{k}|}\,
\left|\bm{J}^{\pm}\cdot\bm{\epsilon}_{\lambda}\right|^2.
\end{equation}
Following the convention given in Ref. \cite{sagh}, we choose the
photon momentum $\bm{k}$ in Eqs.(\ref{cgln+}) and (\ref{cgln-})
along the $z$-axis and take $(\bm{k}\times \bm{q})$ as the
$y$-axis normal to the production plane in the CM frame. In this
coordinate system the current is expressed in terms of the CGLN
amplitudes and given as
$\frac{1}{4}\left|\bm{J}^{\pm}\cdot\bm{\epsilon}_{\lambda}\right|^2
= I^+(\theta)$, where
\begin{eqnarray}\label{differential}
I^+(\theta) &=& |F^+_1|^2+|F^+_2|^2 +2{\rm Re}({F^+_1}^*
F^+_2)\cos\theta \nonumber \\&&+\sin^2\theta \biggl[\frac{1}{2}
(|F^+_3|^2+|F^+_4|^2) +{\rm Re}({F^+_1}^* F^+_4 -{F^+_2}^* F^+_3 +
{F^+_3}^* F^+_4 \cos\theta)\biggr],\nonumber\\ I^-(\theta)
&=&|F^-_1|^2+|F^-_2|^2 +2{\rm Re}({F^{-}_1}^* F^-_2)\cos\theta
\nonumber \\&&+ \sin^2\theta \biggl[\frac{1}{2}
(|F^-_3|^2\sin^2\theta+|F^-_4|^2) -{\rm Re}({F^-_1}^* F^-_3
-{F^-_2}^* F^-_4 + {F^-_2}^* F^-_3\cos\theta)\biggr],
\end{eqnarray}
for each parity of the $\Theta^+(\frac{1}{2}^{\pm})$,
respectively.

\begin{figure}[tbh]
\centering%
\includegraphics[width=10cm]{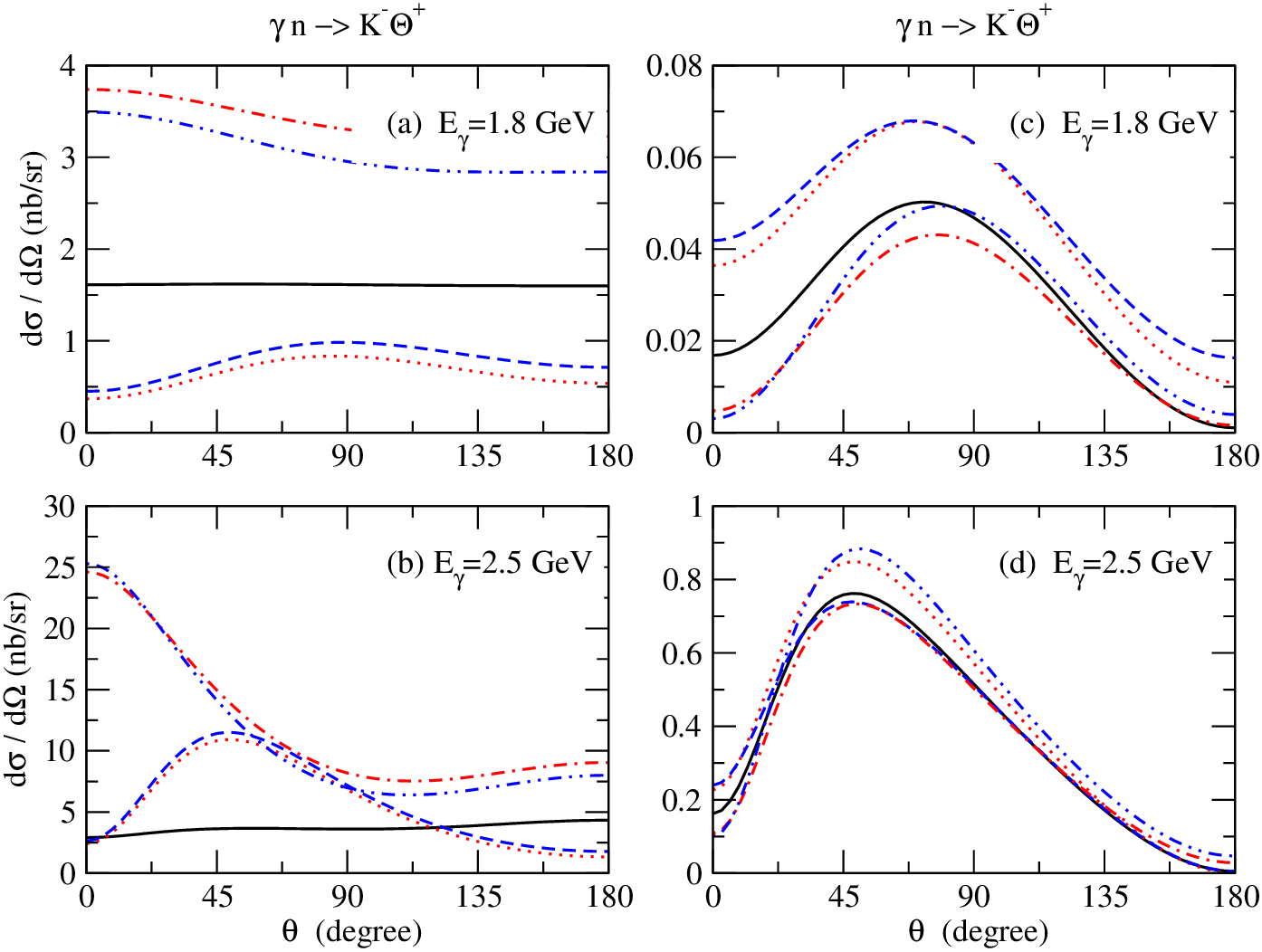}
\caption{(Color online) Angular distributions
$\frac{d\sigma}{d\Omega}$ for $\gamma n \to K^- \Theta^+$ at
$E_\gamma = 1.8$ and $2.5$ GeV; (a), (b) for the
$\Theta^+(\frac{1}{2}^+)$ and (c), (d) for the
$\Theta^+(\frac{1}{2}^-)$. The notations for the curves are the
same with those of Fig. \ref{fig:fig00}.} \label{fig:fig01}
\end{figure}

\begin{figure}[tbh]
\centering%
\includegraphics[width=10cm]{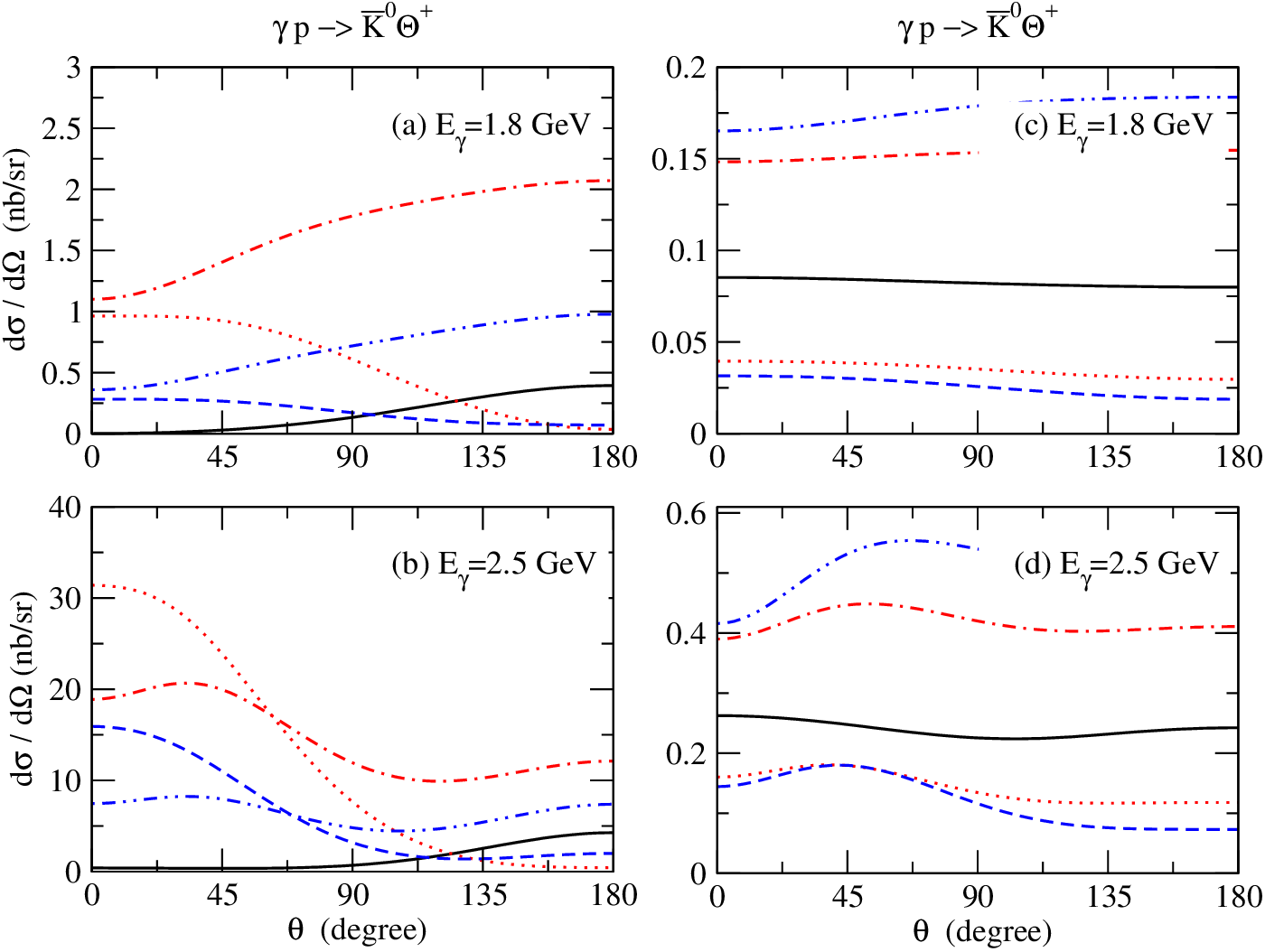}
\caption{(Color online) Angular distributions
$\frac{d\sigma}{d\Omega}$ for $\gamma p\to \bar{K}^{0}\Theta^+$ at
$E_\gamma = 1.8$ and $2.5$ GeV; (a), (b) for the
$\Theta^+(\frac{1}{2}^+)$ and (c), (d) for the
$\Theta^+(\frac{1}{2}^-)$. The notations for the curves are the
same with those of Fig. \ref{fig:fig00}.} \label{fig:fig02}
\end{figure}

Figs. \ref{fig:fig01} and \ref{fig:fig02} show the results
of the angular distribution of Eq.(\ref{diff}) estimated with
coupling constants listed in Table. \ref{cc}. The solid lines in
the figures are the contributions of the Born terms and the
notations for other lines are the same as those of Fig.
\ref{fig:fig00}. In Fig. \ref{fig:fig01} the solid lines of
$\gamma n\to K^-\Theta^+$ near threshold, i.e. at $E_\gamma=$1.8
GeV, show a typical s-wave kaon production of the Born terms for
$\Theta^+(\frac{1}{2}^+)$ and p-wave distribution for
$\Theta^+(\frac{1}{2}^-)$, respectively. These features are quite
distinctive to each other. Note that these are indeed the
consequences already predictable from eq.(\ref{gamma n}),
regardless of any model calculations for the CGLN amplitude
$F^{\pm}_i$ of eq.(\ref{differential}). Furthermore, since such
contrasting features of $\gamma n\to K^-\Theta^+$ are not degraded
by the model-dependence especially due to the $K^*$ and $K_1$
contributions near threshold, $E_{\gamma}=1.8$ GeV,  the angular
distribution near threshold in the case of $\gamma n\to
K^-\Theta^+$ should provide useful informations on the parity of
the $\Theta^+$ without using a polarized photon beam and/or other
polarized observables for that purpose.
For the process $\gamma p\to \bar{K}^0\Theta^+$ in Fig.
\ref{fig:fig02}, however, due to the s-wave nature of the currents
for both parities in eq.(\ref{gamma p}),  the angular
distributions are expected to be isotropic and therefore are not
sensitive to the sign of the $\Theta^+$ parity. Thus, for the case
of $\gamma p\to \bar{K}^0\Theta^+$ we agree with Ref. \cite{reka}
which claims that one should analyze the spin observables to
distinguish between the positive and negative parity of the
$\Theta^+$. These analyses confirm our earlier discussions.

\vspace{0.3cm}

It is natural to extend our analysis to the photon polarization
asymmetry $\Sigma$. The asymmetry of photon polarization is
defined by the difference between the $x$ and $y$ components of
the current in the previous coordinate system. Thus the spin
observable measures an interference between spin-flip and spin
non-flip transitions. From eq.(\ref{gamma n}) different features
depending on the parity of the $\Theta^+$ can be expected in the
measurement of the $\Sigma$  for the $\gamma n\to K^-\Theta^+$
process. However, the case of $\gamma p\to\bar{K}^0\Theta^+$ is
not likely to give a definitive result in the measurement of
$\Sigma$ again as expected from eq. (\ref{gamma p}).

In the same coordinate system as before, the photon polarization
asymmetry is defined by \cite{sagh}
\begin{equation}\label{sigma}
\Sigma=\frac{1}{4
I(\theta)}tr\{J_{y}J_{y}^{\dagger}-J_{x}J_{x}^{\dagger}\,\},
\end{equation}
where the cartesian components of the currents $\bm{J^{\pm}}$ in
this frame are given by
\begin{eqnarray}\label{j+}
&&J_{x}^+=i(F^+_1-F^+_2\cos\theta+F^+_4\sin^2\theta)\sigma_x
+i\sin\theta(F^+_2+F_3^+ +F_4^+\cos\theta)\sigma_z\,,\nonumber\\
&&J_{y}^+=-F_2^+\sin\theta +i(F^+_1-F^+_2\cos\theta)\sigma_y\,,
\end{eqnarray}
for the $\Theta^+(\frac{1}{2}^+)$ and
\begin{eqnarray}\label{j-}
&&J_{x}^-=(F^-_2+F_4^-)\sin\theta +i(F_1^- + F_2^-\cos\theta
-F_3^-\sin^2\theta)\sigma_y\,,\nonumber\\
&&J_{y}^-=-i(F^-_1 + F^-_2\cos\theta)\sigma_x+iF_2^-\sin\theta\sigma_z\,,
\end{eqnarray}
for the $\Theta^+(\frac{1}{2}^-)$, respectively.

\begin{figure}[tbh]
\centering%
\includegraphics[width=10cm]{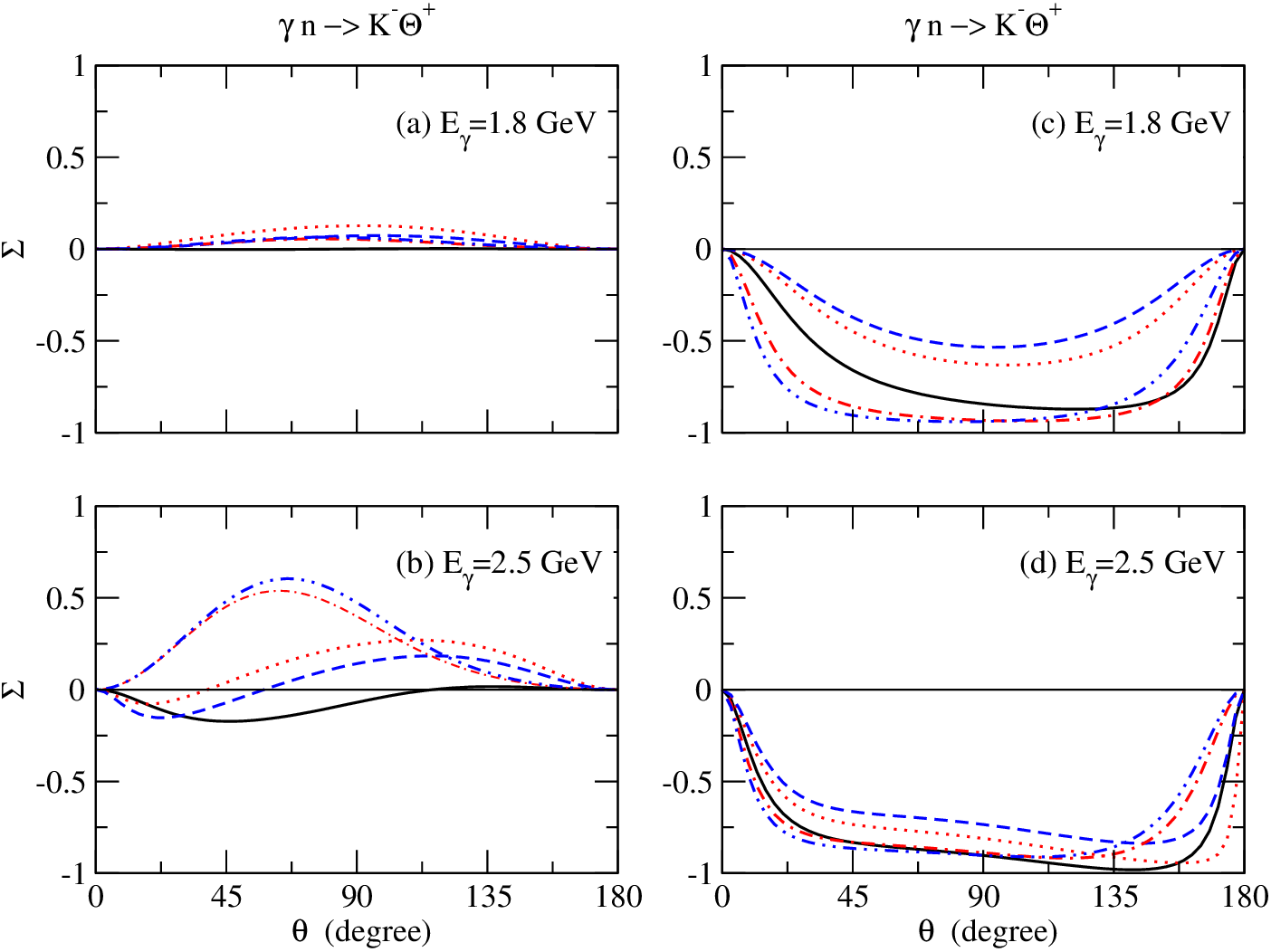}
\caption{(Color online) Polarization asymmetry of photon for
$\gamma n \to K^- \Theta^+$ at $E_\gamma = 1.8$ and $2.5$ GeV;
(a), (b) for the $\Theta^+(\frac{1}{2}^+)$ and (c), (d) for the
$\Theta^+(\frac{1}{2}^-)$.  The notations for the curves are the
same with those of Fig. \ref{fig:fig00}. } \label{fig:fig05}
\end{figure}

\begin{figure}[tbh]
\centering%
\includegraphics[width=10cm]{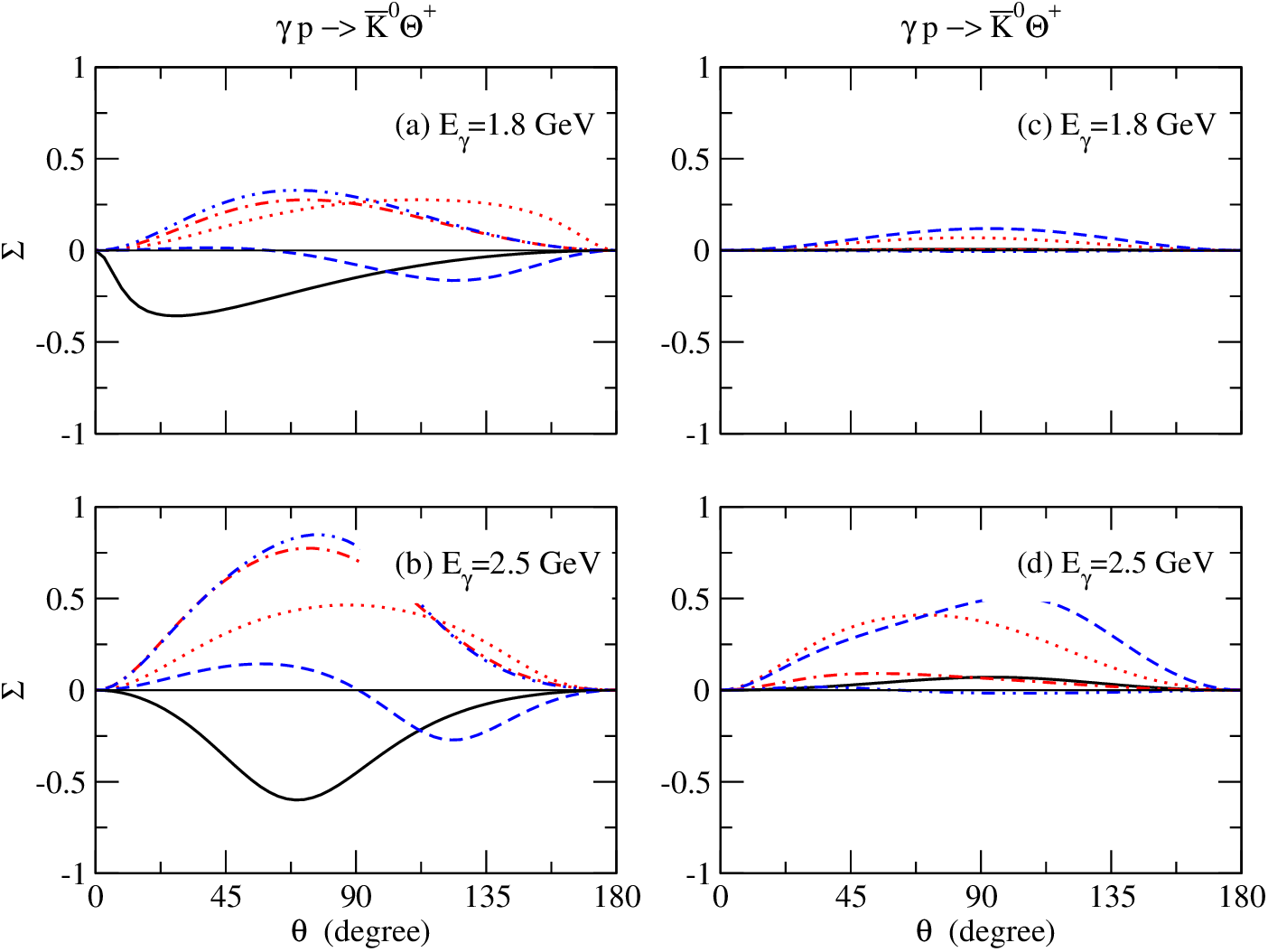}
\caption{(Color online) Polarization asymmetry of photon for
$\gamma p \to \bar{K}^0 \Theta^+$ at $E_\gamma = 1.8$ and $2.5$
GeV; (a), (b) for the $\Theta^+(\frac{1}{2}^+)$ and (c), (d) for
the $\Theta^+(\frac{1}{2}^-)$.  The notations for the curves are
the same with those of Fig. \ref{fig:fig00}. } \label{fig:fig06}
\end{figure}

With normalization by the angular distribution
$I(\theta)=\frac{1}{4}(J_x J_x^{\dagger}+J_y J_y^{\dagger})$, the
results of the photon polarization asymmetry $\Sigma$ are
presented in Fig. \ref{fig:fig05}  for $\gamma n\to K^-\Theta^+$
and Fig. \ref{fig:fig06} for $\gamma p\to \bar{K}^0\Theta^+$,
respectively.
\vspace{0.2cm}
In short, the polarization $\Sigma^{\pm}$ for $\gamma n\to
K^-\Theta^+(\frac{1}{2}^{\pm})$ near threshold can be
characterized by such contrasting features between
$\Sigma^+\approx 0$ and $\Sigma^-\approx -1$. Note that these
contrasting features  are again the consequences of the
conservation of parity and angular momentum plus threshold
kinematics as discussed before. From Eqs. (\ref{j+}) and
(\ref{j-}), the leading order contributions of $F_1^+$ and $F_4^-$
for $\gamma n\to K^-\Theta^+$ near threshold lead to
$J_{x}^+\simeq i F_1^+ \sigma_x$ and $J_{y}^+\simeq i F_1^+
\sigma_y$ for the $\Theta^+(\frac{1}{2}^+)$, and likewise,
$J_{x}^-\simeq F_4^- \sin\theta$, $J_{y}^-\simeq 0$ for the
$\Theta^+(\frac{1}{2}^-)$, respectively. Thus, $\Sigma^+ \approx
0$ can be accounted for by the strong cancellation of the
spin-flip transition, while $\Sigma^- \approx -1$ resulted solely
from the non-flip transition which could arise from the t-channel
kaon exchange in the Born terms. Furthermore, such a distinction
of $\Sigma$ between the positive and negative parity of the
$\Theta^+$ near threshold is not much deteriorated by the
model-dependent contributions of $K^*$ and $K_1$. Therefore,
observation of the photon polarization asymmetry $\Sigma$ for the
process $\gamma n\to K^-\Theta^+$ can also be a useful tool for
determining the parity of the  $\Theta^+$.

For the process $\gamma p\to\bar{K}^0\Theta^+$, however, due to
the dominance of $F_1^+$ and $F_1^-$, the currents in Eqs.
(\ref{j+}) and (\ref{j-}) are approximated as $J_{x}^+\simeq i
F_1^+ \sigma_x$ and $J_{y}^+\simeq i F_1^+ \sigma_y$ for the
$\Theta^+(\frac{1}{2}^+)$ and $J_{x}^-\simeq iF_1^- \sigma_y$,
$J_{y}^-\simeq -i F_1^- \sigma_x$  for $\Theta^+(\frac{1}{2}^-)$,
respectively. Therefore, regardless of the $\Theta^+$ parity, the
polarizations $\Sigma^{\pm}$ are of the s-wave nature close to
threshold and would be $\Sigma^{\pm} \approx 0$. In the case of
the $\gamma p\to \bar{K}^0\Theta^+$ process we, thus, agree with
Ref. \cite{reka} that one needs to measure more polarization
observables beyond the single polarization $\Sigma$ for a
distinction between the $\Theta^+(\frac{1}{2}^+)$ and
$\Theta^+(\frac{1}{2}^-)$.

It is worth to give a few remarks on our result in comparison with
those of Ref. \cite{naka1,zhao01,naka2}. Zhao \cite{zhao01} and
Zhao and Al-Khalili \cite{zhao} also carried out analyses of the
single and double polarization observables of the process $\gamma
n\to K^-\Theta^+$ with polarized photon beams using a quark
potential model and examined kinematical and dynamical aspects for
the purpose of determining the spin and parity of the $\Theta^+$.
Although the model in Ref. \cite{zhao01} is different from ours,
we found that our results are rather close to the results of Ref.
\cite{zhao01} in some cases supporting our findings in this work,
i.e., dominance of the s-wave, and the p-wave nature near
threshold depending on the parity of the $\Theta^+$, respectively.
In particular, our result of the photon polarization $\Sigma$ is
consistent with the corresponding result in Ref. \cite{zhao01}.
Note that the sign convention adopted in Ref. \cite{zhao01} is
opposite to ours due to the redefinition of $\Sigma_A$ to
$\Sigma_W$. Refs. \cite{naka1, naka2} also investigated the photon
polarization $\Sigma$ for the purpose of determining the parity of
the $\Theta^+$. For the model-independent analysis, Ref.
\cite{naka1} exploited the reflection symmetry of the
photoproduction current, which states that the Pauli spin
structure of $J_x^+$ is the same as that of $J_y^-$, and,
likewise, the spin structure of $J_y^+$ as that of $J_x^-$. As can
be seen in Eqs. (\ref{j+}) and (\ref{j-}), our analysis satisfies
the same reflection symmetry that was presented in Ref.
\cite{naka1}. If the reflection symmetry is preserved in the
threshold region, then, the invariance of each component of the
current under an interchange of the parity would make the
determination of the $\Theta^+$ parity obscure just like in the
case of $\gamma p\to\bar{K}^0\Theta^+$ due to the similar shape of
$\Sigma^{\pm}$ between the positive and negative parity. In the
$\gamma p\to\bar{K}^0\Theta^+$ process, the symmetric relation
between the components of $J_{x(y)}^{\pm}$ are preserved and thus
our result of not being able to distinguish the $\Theta^+$ parity
in this process is consistent with the result of Ref.\cite{naka1}.

\subsection{Polarization asymmetries of target nucleon and recoiled $\Theta^+$}

In the previous section, we have demonstrated that  due to the
s-wave nature of both parities in the case of the $\gamma
p\to\bar{K}^0\Theta^+$, it is difficult to determine the parity of
the $\Theta^+$  from the measurement of angular distribution and
photon polarization asymmetry. For a distinction of the $\Theta^+$
parity in the $\gamma p\to \bar{K}^0\Theta^+$ process, let us
consider the polarization asymmetries of the target nucleon and
recoiled $\Theta^+$. We expect that these observables are also
sensitive to the parity of the $\Theta^+$ because the spin and
parity of the initial state is highly correlated with those of the
final state by the conservation laws. For instance, if the
polarization of the $\Theta^+$ in the $\gamma p\to
\bar{K}^0\Theta^+$ process is measured by the strong decay
$\Theta^+\to KN$, similar to the measurement of the $\Lambda^0$
polarization from the $\gamma p\to K^+\Lambda^0$ process in Ref.
\cite{glan}, the decay of outgoing kaon should give different
angular distributions between the two kinds of $\Theta^+$
spin-parity $J^p=1/2^{+}$ and $1/2^{-}$ in accordance with the
partial wave of the $KN$ state either $p$-, or $s$-wave multipole.
Therefore determination of the $\Theta^+$ parity from these
observables could be quite model-independent because such a
correlation is independent of the dynamical details of the
production phenomena. On the other hand, we note that
the nonresonant Born terms except for
the u-channel $\Theta^+$ exchange could give little contribution
since these
observables are given by the imaginary part of the reaction
amplitudes as shown below.
Hence, the features of the observables depend pretty much on how one
handles t-channel $K^*$ and $K_1$ exchanges. The informations on
the coupling constants of these particles are however currently not much
available. With
these in mind, we proceed to figure out how much the observables
could be sensitive to the parity of the $\Theta^+$, in particular
in the case of $\gamma p\to\bar{K}^0\Theta^+$ process.

The polarization asymmetry of the target nucleon is
the measurement of the nucleon spin polarization in the
photoproduction process.  It is defined by
\begin{equation}\label{target0}
T=\frac{d\sigma^+/d\Omega-d\sigma^-/d\Omega}{d\sigma^+/d\Omega
+d\sigma^-/d\Omega}\,,
\end{equation}
where the sign $+(-)$ represents that the spin polarization of the
target nucleon is parallel(antiparallel) to the $y$-axis which is
normal to the plane spanned by the $ \bm{k}\times \bm{q}$ with the
photon momentum $\bm k$ incident to the $z$-axis \cite{sagh}. In
this coordinate system, the target polarization asymmetries
$T^{\pm}$ for each parity of the $\Theta^+$ are given by the
imaginary part of the CGLN amplitudes, i.e.,
\begin{eqnarray}\label{target}
T^+&=&\frac{\sin\theta}{I^+(\theta)}\,{\rm Im}[\,{F_1^+}^* F_3^+
+ {F_2^+}^* F_4^+ +({F_1^+}^* F_4^+ +{F_2^+}^* F^+_3)\cos\theta - {F_3^+}^*
F_4^+\sin^2\theta\,]\,,\nonumber\\
T^-&=&\frac{\sin\theta}{I^-(\theta)} {\rm Im}[\,{F_1^-}^* F_4^-
+{F_2^-}^* F_4^-\cos\theta +({F_2^-}^* F_3^- -{F_3^-}^*
F_4^-)\sin^2\theta\,]\,,
\end{eqnarray}
respectively.
\begin{figure}[tbh]
\centering%
\includegraphics[width=10cm]{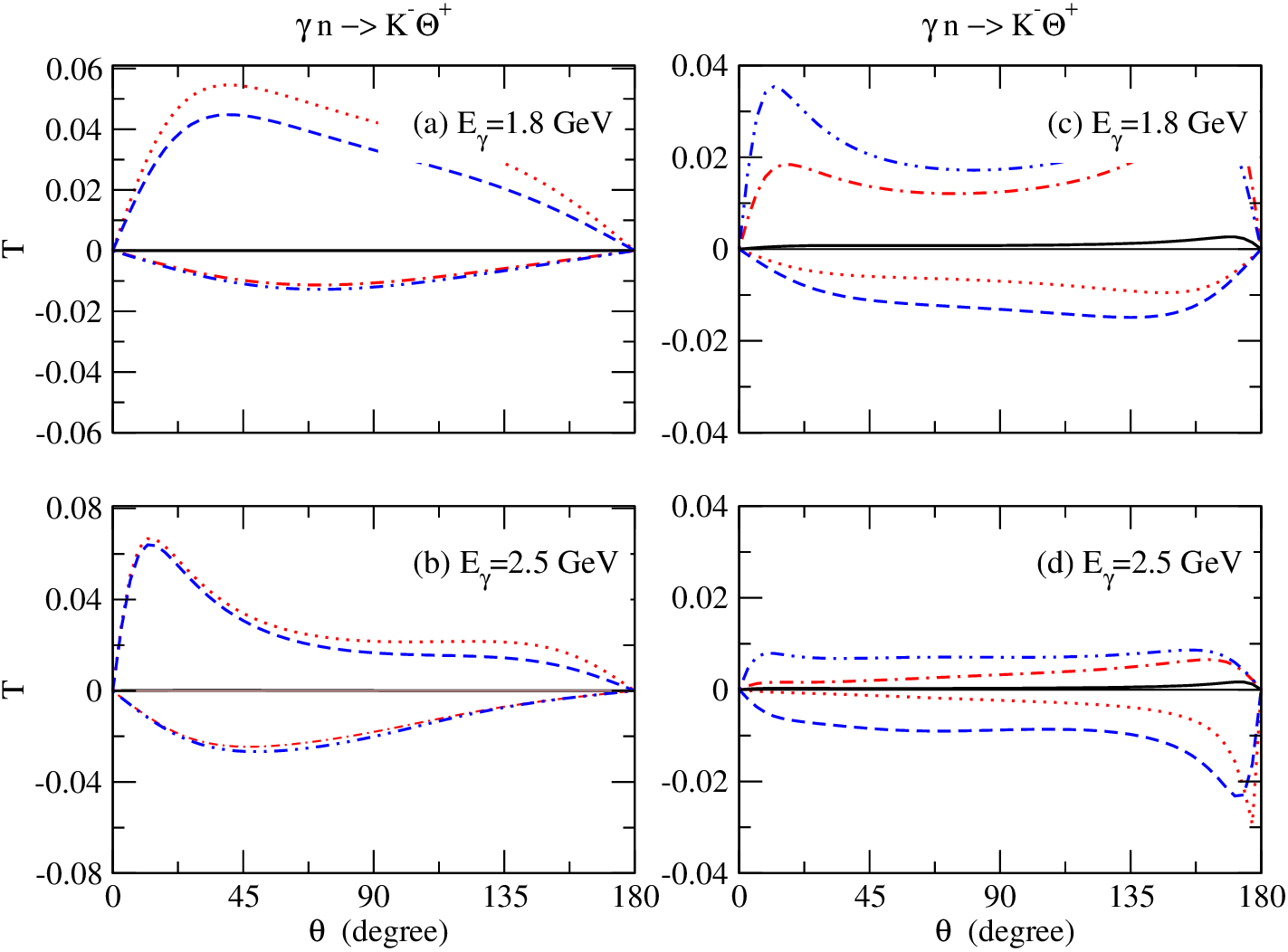}
\caption{(Color online) Polarization asymmetry of target $T$ for
$\gamma n \to K^- \Theta^+$ at $E_\gamma = 1.8$ and $2.5$ GeV;
(a), (b) for the $\Theta^+(\frac{1}{2}^+)$ and (c), (d) for the
$\Theta^+(\frac{1}{2}^-)$. The notations for the curves are the
same with those of Fig. \ref{fig:fig00}. } \label{fig:fig07}
\end{figure}
\begin{figure}[tbh]
\centering%
\includegraphics[width=10cm]{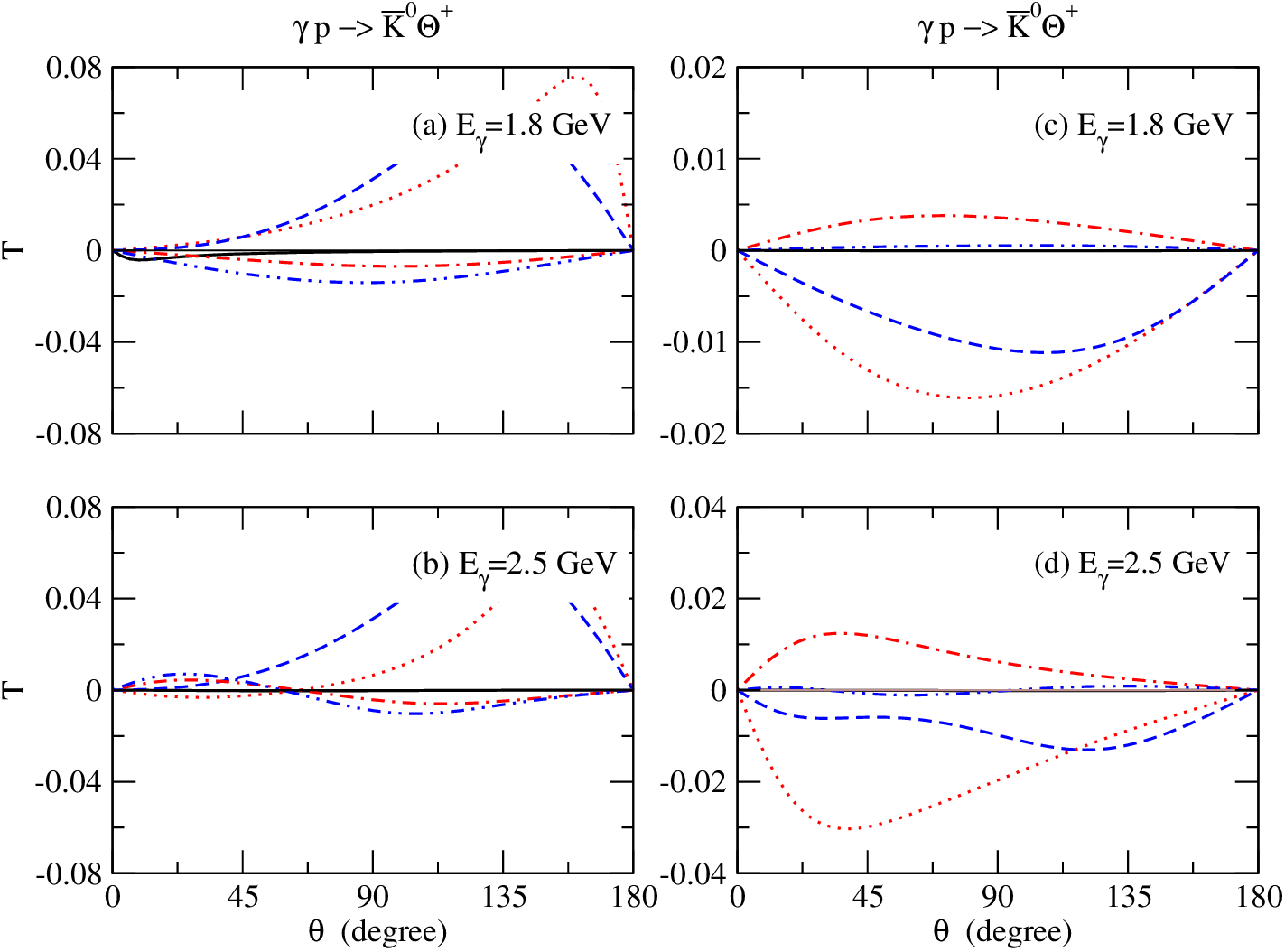}
\caption{(Color online) Polarization asymmetry of target $T$ for
$\gamma p \to \bar{K}^0 \Theta^+$ at $E_\gamma = 1.8$ and $2.5$
GeV; (a), (b) for the $\Theta^+(\frac{1}{2}^+)$ and (c), (d) for
the $\Theta^+(\frac{1}{2}^-)$.  The notations for the curves are
the same with those of Fig. \ref{fig:fig00}. } \label{fig:fig08}
\end{figure}

The polarization asymmetry of the recoiled $\Theta^+$
is defined in the similar fashion, i.e.,
\begin{equation}\label{recoil0}
P=\frac{d\sigma^{(+)'}/d\Omega-d\sigma^{(-)'}/d\Omega}{d\sigma^{(+)'}/d\Omega
+d\sigma^{(-)'}/d\Omega}\,,
\end{equation}
but the sign $(+)'((-)')$ in this case represents that the spin
polarization of the $\Theta^+$ is parallel(antiparallel) to the
$y^{\prime}$-axis normal to the plane spanned by the direction of
$ \bm{k}\times \bm{q}$ with the scattered kaon momentum $\bm q$
directed to the $z^{\prime}$-axis \cite{sagh}. In this coordinate
system, the polarizations $P^{\pm}$ are given by the following
expressions in terms of CGLN amplitude for each parity of the
$\Theta^+$ respectively,
\begin{eqnarray}\label{recoil}
P^+&=&\frac{\sin\theta}{I^+(\theta)} \,{\rm Im}[\,{F^+_1}^*
(2F^+_2 -F^+_3 -F^+_4\cos\theta) 
- {F^+_2}^*(F^+_3\cos\theta\,+F^+_4)+{F^+_3}^* F^+_4\sin^2\theta\,],
\nonumber\\
P^-&=&\frac{\sin\theta}{I^-(\theta)} {\rm Im}[\,2{F^-_1}^* F^-_2
+{F^-_1}^* F^-_4 +{F^-_2}^* F^-_4\cos\theta  
+({F^-_2}^* F^-_3-{F^-_3}^* F^-_4)\sin^2\theta\,].
\end{eqnarray}

\vspace{0.5cm}
\begin{figure}[tbh]
\centering%
\includegraphics[width=10cm]{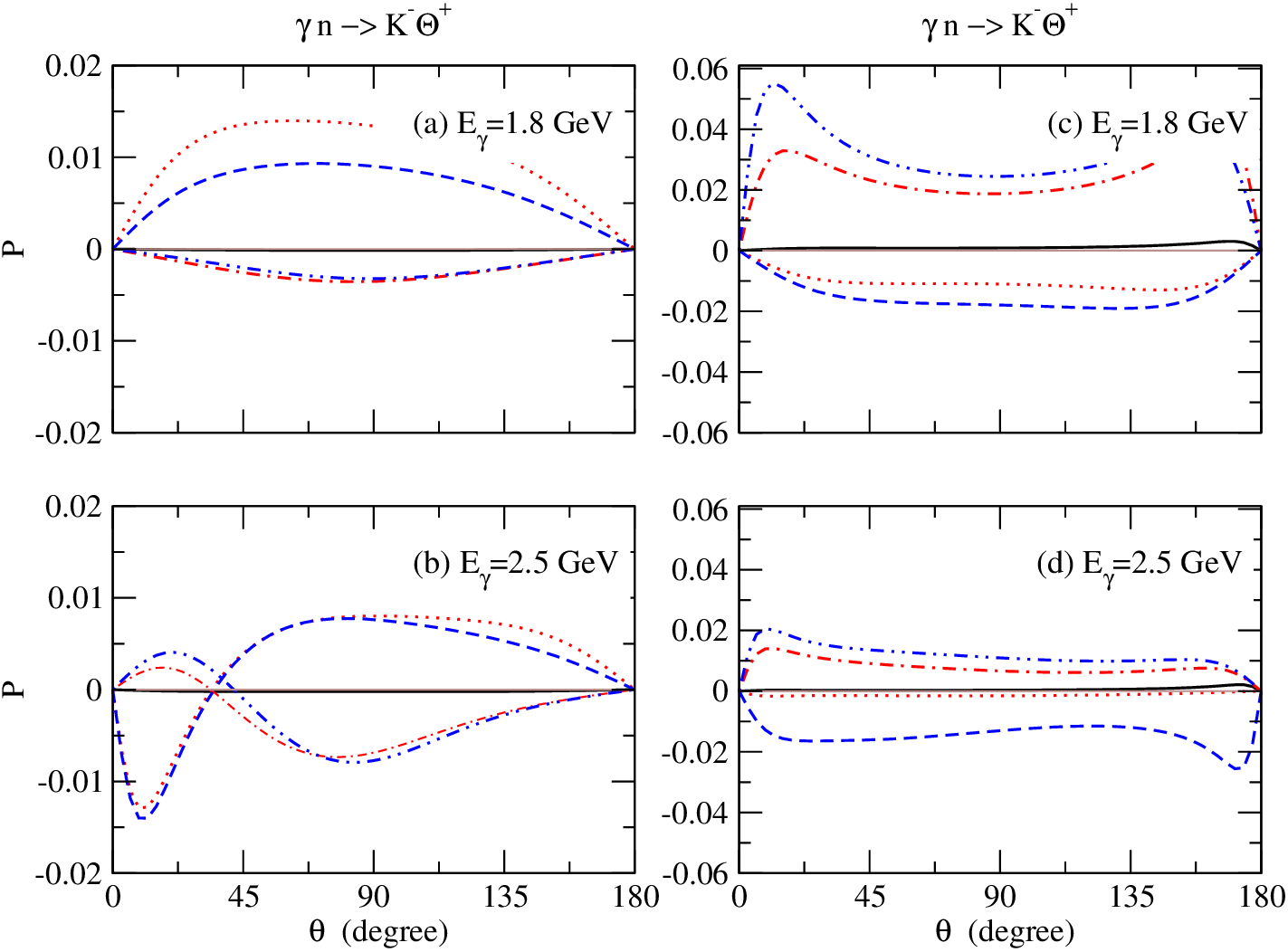}
\caption{(Color online) Polarization asymmetry of recoiled
$\Theta^+$ $P$ for $\gamma n \to K^- \Theta^+$ at $E_\gamma = 1.8$
and $2.5$ GeV; (a), (b) for the $\Theta^+(\frac{1}{2}^+)$ and (c),
(d) for the $\Theta^+(\frac{1}{2}^-)$.  The notations for the
curves are the same with those of Fig. \ref{fig:fig00}.}
\label{fig:fig09}
\end{figure}
\begin{figure}[tbh]
\centering%
\includegraphics[width=10cm]{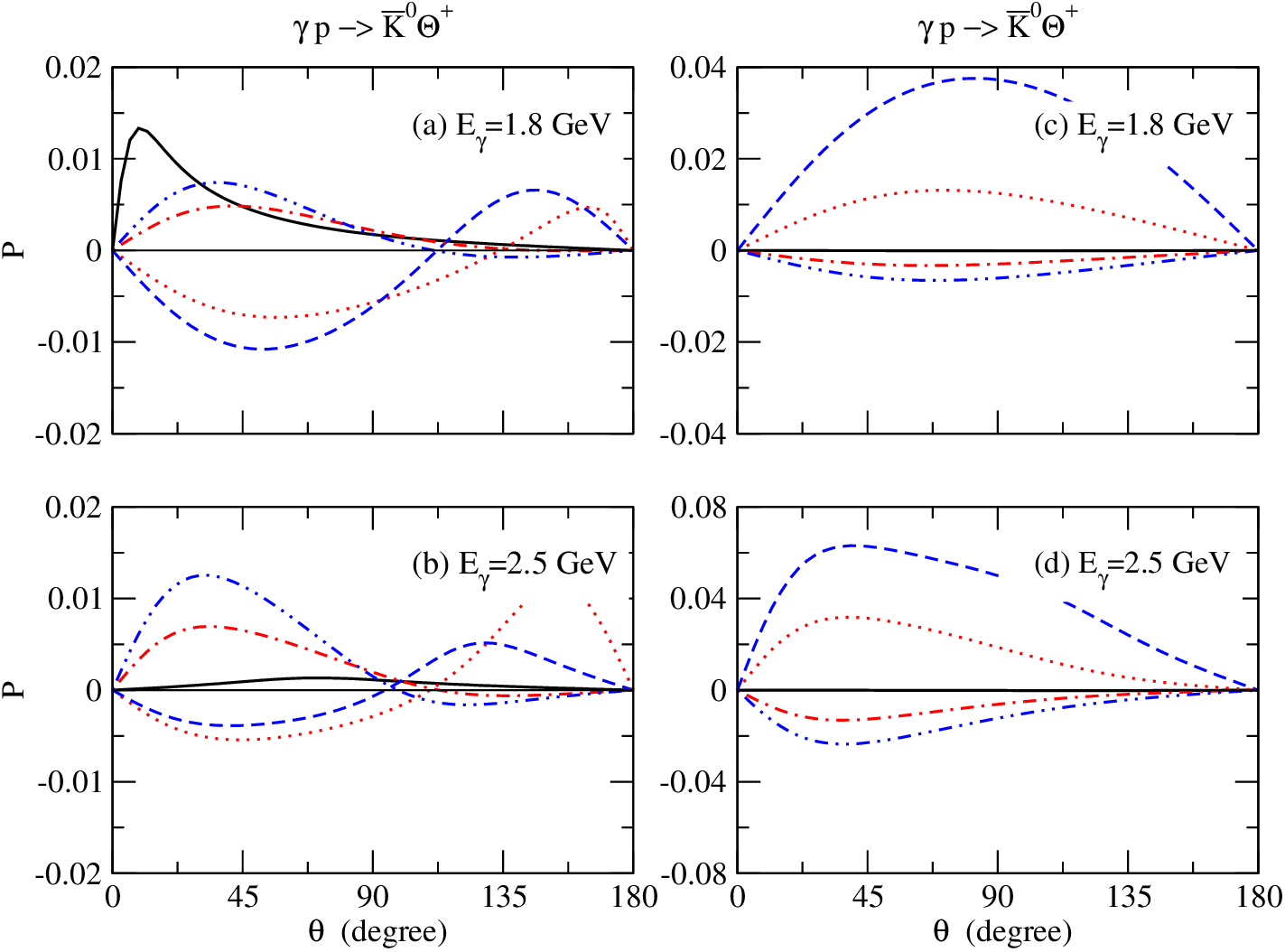}
\caption{(Color online) Polarization asymmetry of recoiled
$\Theta^+$ $P$ for $\gamma p\to \bar{K}^{0}\Theta^+$ at $E_\gamma
= 1.8$ and $2.5$ GeV; (a), (b) for the $\Theta^+(\frac{1}{2}^+)$
and (c), (d) for the $\Theta^+(\frac{1}{2}^-)$. The notations for
the curves are the same with those of Fig. \ref{fig:fig00}.
}\label{fig:fig10}
\end{figure}

We present the results of the target polarization $T$ in Fig.
\ref{fig:fig07} for $\gamma n\to K^-\Theta^+$ and in Fig.
\ref{fig:fig08} for $\gamma p\to \bar{K^0}\Theta^+$, respectively.
For the polarization asymmetry $P$, we show the results in Figs.
\ref{fig:fig09} and \ref{fig:fig10} for $\gamma n\to K^-\Theta^+$
and $\gamma p\to \bar{K^0}\Theta^+$, respectively.
In these figures, we see that the given expressions in Eqs.
(\ref{target}) and (\ref{recoil}) provide general rules for $T$
and $P$ similar to $ \Sigma$, i.e., they are identical to  zero at
$\theta=0^{\circ}$ and $180^{\circ}$ due to the dependence on the
gross factor of $\sin\theta$. Moreover, as the CGLN amplitudes of
the $K^*$ and $K_1$ given by eqs. (\ref{cglnftn+}) and
(\ref{born+}) do not depend on the angle significantly near
threshold, the features of these observables at $E_{\gamma}=1.8$
GeV exhibit the predominating $\sin\theta$ behavior with their
signs reversed depending on the signs of the $K^*$ and $K_1$
coupling constants. As shown in Figs. \ref{fig:fig07} and
\ref{fig:fig08}, it is interesting to see that the sign of target
polarization $T$ becomes opposite to each other whenever the
parity of $\Theta^+$ is changed, and this feature is common to
both processes $\gamma n\to K^-\Theta^+$ and $\gamma
p\to\bar{K}^0\Theta^+$. Furthermore, as shown in Figs.
\ref{fig:fig09} and \ref{fig:fig10}, the $\Theta^+$ polarization
$P$ also shows the pattern very similar to the case of
polarization $T$. Thus, we find that the observables $T$ and $P$
are highly sensitive to the $\Theta^+$ parity.
\vspace{0.5 cm}
\begin{figure}[tbh]
\centering%
\includegraphics[width=10cm]{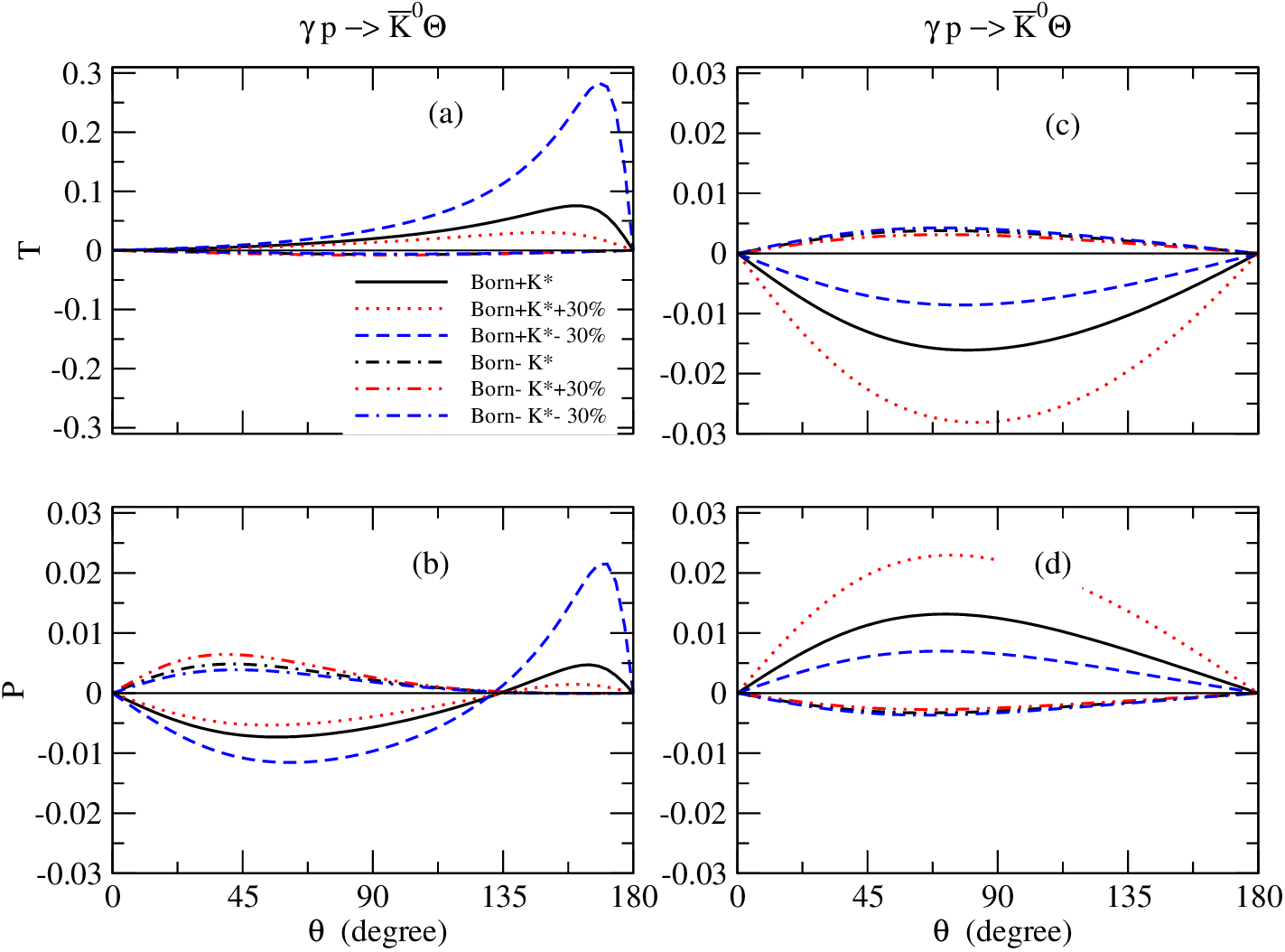}
\caption{(Color online) Dependence of polarization asymmetries  T
and P on the coupling constant $g_{K^*N\Theta}$ for $\gamma p \to
\bar{K}^0 \Theta^+$ at $E_\gamma = 1.8$ GeV; (a), (b) for the
$\Theta^+(\frac{1}{2}^+)$ and (c), (d) for the
$\Theta^+(\frac{1}{2}^-)$. $g_{KN\Theta}$ is fixed in any case as
given in Table \ref{cc} and contribution of $K_1$ is neglected for
simplicity.  Note that the notations for the curves are completely
different from those in Figs. \ref{fig:fig07} -  \ref{fig:fig10}.
The solid lines(the dotted lines of Figs. \ref{fig:fig08} and
\ref{fig:fig10}) are the sum of the $\Theta^+$ pole term and $K^*$
exchange with coupling constants given in Table \ref{cc}. Other
lines correspond to the cases when the $K^*$ coupling constant in
the solid line increases, or decrease its value by $30 \%$ as
denoted by the legend in this figure. } \label{fig:fig11}
\end{figure}
\begin{figure}[tbh]
\centering%
\includegraphics[width=10cm]{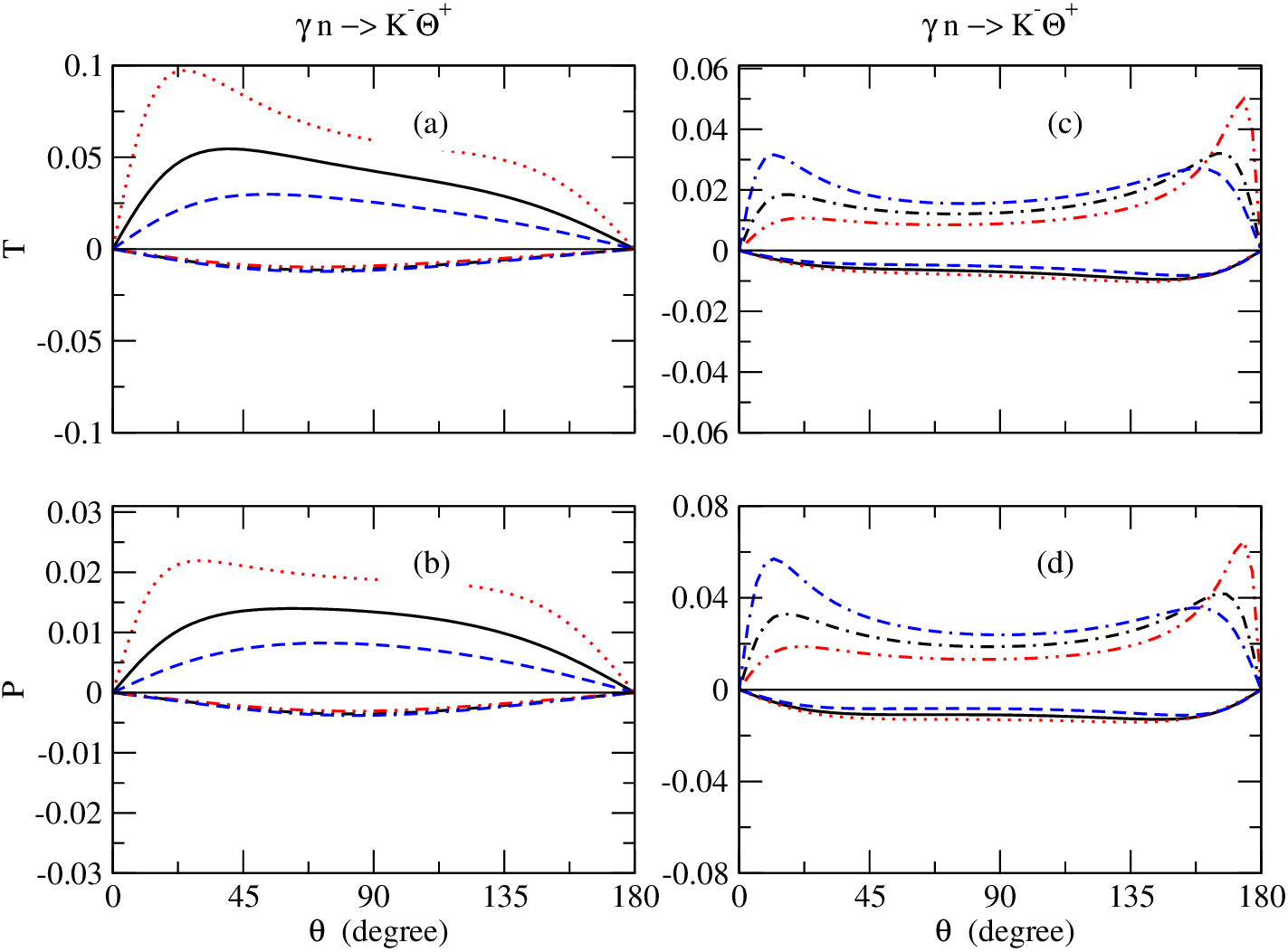}
\caption{(Color online) Dependence of polarization asymmetries T
and P on the coupling constant $g_{K^*N\Theta}$ for $\gamma n \to
K^- \Theta^+$ at $E_\gamma = 1.8$ GeV; (a), (b) for the
$\Theta^+(\frac{1}{2}^+)$ and (c), (d) for the
$\Theta^+(\frac{1}{2}^-)$. $g_{KN\Theta}$ is fixed in any case as
given in Table \ref{cc} and contribution of $K_1$ is neglected for
simplicity. The solid lines(the dotted lines of Figs.
\ref{fig:fig07} and \ref{fig:fig09}) are the sum of the $\Theta^+$
pole term and $K^*$ exchange with coupling constants given in
Table \ref{cc}. Other lines correspond to the cases when the $K^*$
coupling constant in the solid line increases, or decrease its
value by $30 \%$ as denoted by the legend in Fig. \ref{fig:fig11}
above. The notations for the curves are the same with those of
Fig. \ref{fig:fig11}}\label{fig:fig12}
\end{figure}
In addition we suppose that the features of the spin observables
aforementioned would be less dependent on the $K^*$ and $K_1$
coupling constants near threshold, because they are, by
definition, given by  the ratios normalized by differential cross
section. This point should be clarified, however.
In order to see the model-dependence of the $T$ and $P$ on the
coupling constants of vector mesons, we reproduce these
observables by changing the coupling constants $g_{K^*N\Theta}$,
and $g_{K_1 N\Theta}$ by an amount of $30 \%\,$ increase or
decrease. The results are shown in Figs. \ref{fig:fig11} and
\ref{fig:fig12}. As a reference, the solid line is presented in
the figure with coupling constants $g_{KN\Theta}=0.984(0.137)$,
$g_{K^*N\Theta}=1.704(0.08)$ fixed from Table \ref{cc}. Other
lines correspond to the cases when the coupling constant of the
$K^*$ is changed by $30 \%$ larger, or smaller than the solid
line. Since the role of $K_1$ by such a change is not significant,
the contribution of $K_1$ is neglected in the figures for
simplicity. As can be seen in the figures, the change of $K^*$
coupling constant will not alter the contrasting features between
the positive and negative parity of the $\Theta^+$. Thus, the
decisive features of the spin observables $T$ and $P$ are not
affected significantly by the uncertainty of the $g_{K^*N\Theta}$.
Conclusively, the measurement of spin observables $T$ and $P$ can
help determine the parity of the $\Theta^+$ in the future
experiment especially from the $\gamma p\to \bar{K}^0\Theta^+$
process, which is more accessible than the neutron target.

\section{Summary and discussion}

In this work, we have analyzed the unpolarized angular
distribution and single polarization observables for the $\gamma
N\to K\Theta^+$ process. Our analysis illustrates that the
amplitude of $\gamma n\to K^-\Theta^+$ near threshold is
characterized by the dominance of the s-wave (p-wave) multipole
for the positive (negative) parity of the $\Theta^+$. For the
$\gamma n\to K^-\Theta^+$ process, therefore, measurements of
unpolarized angular distribution and photon polarization asymmetry
can be utilized for the determination of the $\Theta^+$ parity.
However, the situation becomes quite different in the case of
$\gamma p\to \bar{K}^0\Theta^+$ process because the direct
coupling of photon to neutral $\bar{K}^0$ meson is absent. Due to
the s-wave nature common to both parities of the $\Theta^+$,
measurements of such observables for $\gamma p\to
\bar{K}^0\Theta^+$ near threshold are not so effective to reveal a
distinction between the two opposite parities of the $\Theta^+$.
As usual in physical processes, the symmetry and the dynamics are
strongly correlated to each other in the case of these
observables. Taking into account the threshold kinematics of CGLN
amplitude, we further confirm that the features of angular
distribution and photon polarization asymmetry near threshold are
natural consequences of the first principle conservation laws of
parity and angular momentum. For determination of the $\Theta^+$
parity using the $\gamma p\to\bar{K}^0\Theta^+$ process, we
investigated the polarization asymmetries of target and recoiled
$\Theta^+$ and found out that these observables could serve
for such purpose. The measurement of the
spin-dependent observables near threshold should be
encouraged, with special emphasis placed on the conservation laws
and the threshold kinematics for model-independent determination of the
$\Theta^+$ parity.

\acknowledgments

This work was supported
by a grant from the U.S.
Department of Energy (DE-FG02-96ER 40947). CRJ thanks to the hospitality
provided by the Department of Physics at Seoul
National University where he took a sabbatical leave for the spring
semester of the year 2005 and completed this work.

\end{document}